\documentclass[twocolumn,superscriptaddress,pra,footinbib,notitlepage]{revtex4-1}
\usepackage[dvipdfmx]{graphicx}
\usepackage{color}
\usepackage{epsfig}
\usepackage{amsmath,amssymb,latexsym}
\usepackage{ascmac}
\usepackage{enumitem}
\usepackage{comment}
\usepackage{braket}
\usepackage{theorem}
\newcommand{\ketbra}[2]{| #1 \rangle \langle #2 |}

\newcommand{\tr}[1]{\mathrm{Tr}\left(#1\right)}

\newcommand{\cpr}[2]{\mathrm{Pr}\left[#1\middle|#2\right]}
\newcommand{\abs}[1]{\lvert#1\rvert}
\newcommand{\relmiddle}[1]{\mathrel{}\middle#1\mathrel{}}

\newtheorem{theo}{Theorem}
\newtheorem{lem}{Lemma}

\newtheorem{fact}{Fact}

\begin{document}
\title{
Differential-phase-shift quantum key distribution protocol with small number of random delays
}
\author{Yuki Hatakeyama}
\affiliation{Graduate School of Engineering Science, Osaka University,
Toyonaka, Osaka 560-8531, Japan}
\author{Akihiro Mizutani}
\email{mizutani@qi.mp.es.osaka-u.ac.jp}
\affiliation{Graduate School of Engineering Science, Osaka University,
  Toyonaka, Osaka 560-8531, Japan}
\author{Go Kato}
\affiliation{
NTT Communication Science Laboratories, NTT Corporation, 3-1 Morinosato Wakamiya, Atsugi, Kanagawa 243-0198, Japan}
\author{Nobuyuki Imoto}
\affiliation{Graduate School of Engineering Science, Osaka University,
  Toyonaka, Osaka 560-8531, Japan}
\author{Kiyoshi Tamaki}
\affiliation{
NTT Basic Research Laboratories, NTT Corporation, 3-1 Morinosato Wakamiya, Atsugi, Kanagawa 243-0198, Japan}
\begin{abstract}

The differential-phase-shift (DPS) quantum key distribution (QKD) protocol was proposed aiming at simple implementation, but it can tolerate only a small disturbance in a quantum channel.
The round-robin DPS (RRDPS) protocol could be a good solution for this problem, which in fact can tolerate even up to $50\%$ of a bit error rate.
Unfortunately, however, such a high tolerance can be achieved only when we compromise the simplicity, \textit{i.e.}, Bob's measurement must involve a large number of random delays ($\abs{\mathcal{R}}$ denotes its number), and in a practical regime of $\abs{\mathcal{R}}$ being small, the tolerance is low.
In this paper, we propose a new DPS protocol to achieve a higher tolerance than the one in the original DPS protocol, in which the measurement setup is less demanding than the one of the RRDPS protocol for the high tolerance regime.
We call the new protocol the small-number-random DPS (SNRDPS) protocol, and in this protocol, we add only a small amount of randomness to the original DPS protocol, \textit{i.e.}, $2\leq\abs{\mathcal{R}}\leq10$.
In fact, we found that the performance of the SNRDPS protocol is significantly enhanced over the original DPS protocol only by employing a few additional delays such as $\abs{\mathcal{R}}=2$.
Also, we found that the key generation rate of the SNRDPS protocol outperforms the RRDPS protocol without monitoring the bit error rate when it is less than $5\%$ and $\abs{\mathcal{R}}\leq10$.
Our protocol is an intermediate protocol between the original DPS protocol and the RRDPS protocol, and it increases the variety of the DPS-type protocols with quantified security.

\end{abstract}
\maketitle
\section{introduction}
Quantum key distribution (QKD) holds promise for realizing information-theoretically secure communication between two distant parties (Alice and Bob) against any eavesdropper (Eve). 
Since the first invention of the BB84 protocol~\cite{bennett1984advances}, many QKD protocols have been proposed~\cite{ekert1991quantum, bennett1992quantum, bruss1998optimal, inoue2002differential, scarani2004quantum, gisin2004towards}.
Among them, the differential-phase-shift (DPS) QKD~\cite{inoue2002differential} can be rather simply implemented with a passive detection unit.
A field demonstration of the DPS protocol~\cite{sasaki2011field} has been already been conducted, and the information-theoretical security proof of the DPS protocol has been established by Tamaki \textit{et al}~\cite{tamaki2012unconditional, Note1}.
Unfortunately, however, this proof shows that the DPS protocol can tolerate only a small bit error rate regime (less than $4\%$ with a typical block length of $L$ light pulses, say $L=32$).

Recently, in order to solve this problem, 
a new type of protocol called the round-robin differential-phase-shift (RRDPS) QKD protocol~\cite{sasaki2014practical} was proposed. 
This is a modified protocol from the original DPS protocol in that Bob's measurement has a freedom to randomly choose which pair of the incoming pulses to be interfered.
This modification brings a distinct feature to the RRDPS protocol that the security can be guaranteed without monitoring any disturbance between Alice and Bob.
Moreover, when the number of random delays (we denote it by $\abs{\mathcal{R}}$) is large, the RRDPS protocol has a strong tolerance to the bit error rate, and surprisingly it can tolerate the bit error rate of even $50\%$ when $\abs{\mathcal{R}}\to\infty$.
Thanks to these features, the RRDPS protocol has attracted theoretical works~\cite{mizutani2015robustness, zhang2015round, takesue2015experimental, chau2016qudit, sasaki2016quantum}, and proof-of-principle experiments 
have been demonstrated~\cite{guan2015experimental, li2016experimental, takesue2015experimental, wang2015experimental}. 
Unfortunately, however, an experimental implementation of the RRDPS protocol is not as simple as the one of the original DPS protocol. 
One of the main technological challenges for its realization is to switch the delay at random for each block of the pulses at Bob's measurement.
According to the number of the random delays, some passive interferometers~\cite{takesue2015experimental} or a variable-delay interferometer~\cite{li2016experimental} or some optical switches~\cite{wang2015experimental} are needed for Bob's measurement.
Obviously, when the number of the random delays increases, an implementation of Bob's measurement will be complicated.
Hence, from a practical viewpoint, it is preferable to implement the RRDPS protocol with small $\abs{\mathcal{R}}$, for instance, $\abs{\mathcal{R}}=4$ as demonstrated in~\cite{takesue2015experimental}.

In this paper, we consider to improve the bit error tolerance of the DPS protocol without significantly increasing the difficulties of its implementation.
For this, we consider to add only a small amount of randomness, say $2\leq\abs{\mathcal{R}}\leq10$, to the DPS protocol.
This modification can also be seen as the modification from the RRDPS protocol in that the new protocol exploits more pulses than that of the RRDPS protocol for a given $\abs{\mathcal{R}}$.
Importantly, this modification does not increase any experimental difficulty at Bob's side.
We call the new protocol the small-number-random DPS (SNRDPS) protocol.
We present the information-theoretical security of our protocol, in which we have made some assumptions on the devices. 
In particular, we assume perfect phase modulations (namely, Alice's phase modulation is exactly 0 or $\pi$)
and block-wise phase randomization (the state of the $L$ pulses is a classical mixture of photon number states).
With these assumptions, we prove the security based on the Shor-Preskill's security proof~\cite{shor2000simple}. 
By using the result of the security proof, we compare the performance of the SNRDPS protocol and the one of the original DPS protocol.
As a result, we found that the key generation rate is significantly improved only with a few additional delays, say $\abs{\mathcal{R}}=2$.
For instance, if the bit error rate $e^{(\textrm{b})}$ is $2\%$, the key generation rate of the SNRDPS protocol with $\abs{\mathcal{R}}=2$ scales as $\mathcal{O}(\eta^{3/2})$ with a channel transmittance $\eta$ in a longer distance regime while the original DPS protocol scales as $\mathcal{O}(\eta^2)$.
Also, when $e^{(\textrm{b})}=5\%$, the SNRDPS protocol with $\abs{\mathcal{R}}=2$ gives a positive key generation rate while the original DPS protocol cannot give a positive one. 
Therefore, the small number of random delays provides a significant improvement in the resulting key generation rate compared to the one of the DPS protocol.

Moreover, we compare the key generation rates of the SNRDPS protocol and the ones of the RRDPS protocol without monitoring the disturbance when the same number of the random delays $\abs{\mathcal{R}}$ is employed between two protocols.
Consequently, we found that the SNRDPS protocol gives a better key generation rate than the one of the RRDPS protocol without monitoring the bit error rate when $\abs{\mathcal{R}}$ is less than 10 and the bit error rate is small such as less than $5\%$.

This paper is organized as follows.
First, in Sec.~\ref{sec:DPS-type-QKD} we explain the DPS-type QKD protocol including the assumptions on Alice and Bob's devices.
Next, in Sec.~\ref{sec:security-proof} we prove the security of the DPS-type protocol, where our security proof is based on the Shor-Preskill's security proof~\cite{shor2000simple}.
After that, in Sec.~\ref{sec:key-generation-rates} we show the simulation results for the SNRDPS protocol, and compare the key generation rates with the various numbers of the random delays $\abs{\mathcal{R}}=\{2, 4, 6, 8, 10\}$.
Finally, we summarize the paper in Sec.~\ref{sec:conclusion}. 

\section{DPS-type QKD protocol}
\label{sec:DPS-type-QKD}
\begin{figure*}[t]
\begin{center}
\includegraphics[width=15cm]{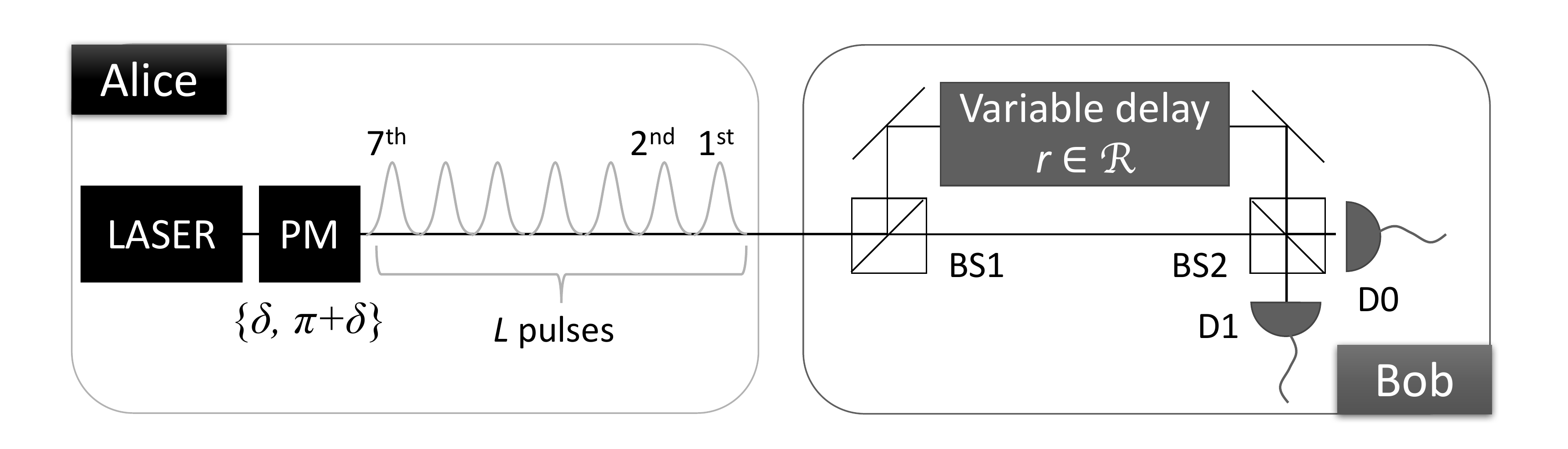}
\caption{Schematics of the SNRDPS protocol for $L=7$ and $\mathcal{R}=\{1,6,2,5\}$. Alice sends 7 coherent pulses after she applies a random phase shift $\delta$ or $\delta+\pi$ to each of the pulses with a phase modulator (PM), where $\delta$ is chosen uniformly and at random from $[0, 2\pi)$. After Bob receives the incoming 7 pulses, he splits them into two blocks of 7 pulses with the first beam splitter (BS1), and shifts backward by $rT$ to one of the blocks. Here, $T$ denotes the interval between two adjacent pulses of the incoming block. By using the second beam splitter (BS2) and two photon-number-resolving detectors (D0 and D1), he observes the relative phase of two pulses in the block.
Note that, each bit in the sifted key is generated from a block where Bob has detected one photon from one pair of the interfering pulses and has detected the vacuum in all the other pulses.
We call such an event \textit{detected} event.
For example, if $r=2$ is chosen and Bob detects exactly one photon in the pair of ($3^\mathrm{rd}$, $5^\mathrm{th}$)-interfering pulses and detects the vacuum in all of the other pulses, he obtains the relative phase between the $3^\mathrm{rd}$ and the $5^\mathrm{th}$ interfering pulses.
If the relative phase is $0~(\pi)$, he obtains the sifted bit $s_\textrm{B}=0~(1)$.}
\label{fig:actual_scheme}
\end{center}
\end{figure*}

In this section, before providing the description of the actual protocol, we first list up the assumptions on Alice and Bob's devices.
See FIG.~\ref{fig:actual_scheme} for the actual setup. 

\subsection{Assumptions on Alice and Bob's devices}
First, we describe the assumptions on Alice's source. We assume that it emits a single-mode coherent light pulse, and Alice splits its pulse into a block of $L$ pulses.
The $L$ pulses are block-wise phase randomized, namely, the quantum state of the $L$ pulses is described as a classical mixture of photon number states.
The relative phase between the adjacent pulses is modulated by 0 or $\pi$ according to her randomly chosen bit 0 or 1, respectively.

Next, as for Bob's device, it is equipped with two photon-number-resolving (PNR) detectors that can discriminate among 0, 1, and more than 1 photon.
He first splits $L$ incoming pulses into two blocks of $L$ pulses by using a $50:50$ beam splitter (BS), shifts backward only one of the $L$-pulse blocks by $r$ that is chosen randomly from the set $\mathcal{R}\subset\{1, 2, \dots, L-1\}$.
Then, the first $L-r$ pulses in the shifted block will be interfered with the last $L-r$ pulses in the other block with another $50:50$ BS, and then Bob performs a photon measurement with the PNR detectors.
Each of the detectors corresponds to the bit value of 0 and 1, respectively (see FIG.~\ref{fig:actual_scheme}).
Finally, we assume that there is no side-channel.
\subsection{DPS-type QKD}
We describe ``DPS-type" QKD protocol, which is the generalization of the DPS QKD protocol in that it employs the arbitrary number of random delays, and therefore DPS-type protocol includes both the original DPS protocol and the RRDPS protocol.
The protocol of the DPS-type QKD runs as follows.
\begin{enumerate}[label=(A\arabic*)]
\setlength{\parskip}{0cm}
\setlength{\itemsep}{0cm}
\item Alice generates a random $L$-bit string $\vec{s}\equiv(s_1, s_2, \dots, s_L)$, a random number $\delta\in[0, 2\pi)$, and then she prepares a block of $L$ coherent pulses in the following state
\begin{align}
\ket{\Psi}=\bigotimes_{k=1}^L\ket{(-1)^{s_k}\mathrm{e}^{\mathrm{i}\delta}\alpha}_k,
\end{align}
where $\ket{(-1)^{s_k}\mathrm{e}^{\mathrm{i}\delta}\alpha}_k$ represents the coherent state of the $k^\mathrm{th}$ pulse.
She sends $\ket{\Psi}$ to Bob through a quantum channel.
\item Bob splits the incoming $L$ pulses into two $L$-pulse blocks by using the $50:50$ BS.
He applies a delay $rT$ to one of the paths in the Mach-Zehnder interferometer, where $T$ denotes the interval between two adjacent pulses in the block and $r$ is chosen uniformly at random from the set $\mathcal{R}\subset\{1, 2, \dots, L-1\}$.
After that, Bob makes interference between two $L$-pulse blocks by using the other $50:50$ BS and performs the photon detection with the photon number resolving detectors.
Let us call the event \textit{detected} if he detects exactly one photon in the pair of ($k^\mathrm{th}$, $(k+r)^\mathrm{th}$) ($1\leq k\leq L-r$) interfering pulses and detects the vacuum in all of the other pulses (including $2r$ half pulses that do not interfere with any other half pulses). If the event is not \textit{detected}, Alice and Bob skip steps (A3) and (A4).
\item Bob takes note of the detected bit value $s_\textrm{B}$ and announces the pair of numbers $(i, j)=(k, k+r)$ over an authenticated public channel.
\item Alice takes note of the bit value $s_\mathrm{A}=s_k\oplus s_{k+r}$.
\item Alice and Bob repeat steps (A1) through (A4) $N$ times, and let $NQ$ be the number of the \textit{detected} events. 
\item Alice and Bob randomly select a small portion $\xi$ of $NQ$ \textit{detected} events, and compare the bit values over an authenticated public channel. This gives the estimate of the bit error rate. 
\item Alice and Bob discuss over an authenticated public channel to perform error correction and privacy amplification on the remaining portion to share a final key of length $GN(1-\xi)$.
\end{enumerate}
Note that the DPS-type protocol includes the original DPS and the RRDPS protocols by choosing $\mathcal{R}$ in step (A2) as $\mathcal{R}=\{1\}$ and $\mathcal{R}=\{1, 2, \dots, L-1\}$, respectively. 
Also, we define the SNRDPS protocol by setting $\mathcal{R}=\bigcup_{m=1}^{t}\{m, L-m\}$ with $0<t<L/2$.

\section{Security proof}
\label{sec:security-proof}
In this section, we prove the security of the SNRDPS protocol with $\mathcal{R}=\bigcup_{m=1}^{t}\{m, L-m\}$ for $0<t<L/2$. Our security proof can be summarized as follows. 
First, in Sec.~\ref{subsec:dial-measurement} we convert the actual protocol to an \textit{alternative} protocol for simplicity of the analysis, where Bob performs the alternative measurement (we call it the dial measurement) instead of the actual one. 
Note that, by switching Bob's actual measurement with the delays $r$ and $L-r$ uniformly at random, we show that he can simulate the dial measurement characterized by the delay $r$ with a $50\%$ of additional detection loss (see Lemma~\ref{fact:dial-actual-relation} below).
Therefore, by introducing the additional loss in the alternative measurement, the dial measurement is equivalent to the actual measurement, and therefore we can employ the alternative measurement in the security proof.
Next, in Sec.~\ref{subsec:virtual-protocol} we introduce an entanglement distillation protocol as a \textit{virtual} protocol to prove the security of the protocol with the alternative measurement.
After that, we construct the POVM elements corresponding to the bit and phase error rates in Sec.~\ref{subsec:POVM-elements}, and derive a relation between the bit and phase errors by employing some constraint on Alice's sending state in the virtual protocol in Sec.~\ref{subsec:relation-bit-phase-errors}, and obtain an upper bound on the phase error rate as the function of the bit error rate in Sec.~\ref{subsec:evaluation-omega}.

\begin{figure*}[!t]
\begin{center}
\begin{tabular}{c}

\begin{minipage}{0.5\hsize}
\begin{center}
\includegraphics[width=8cm]{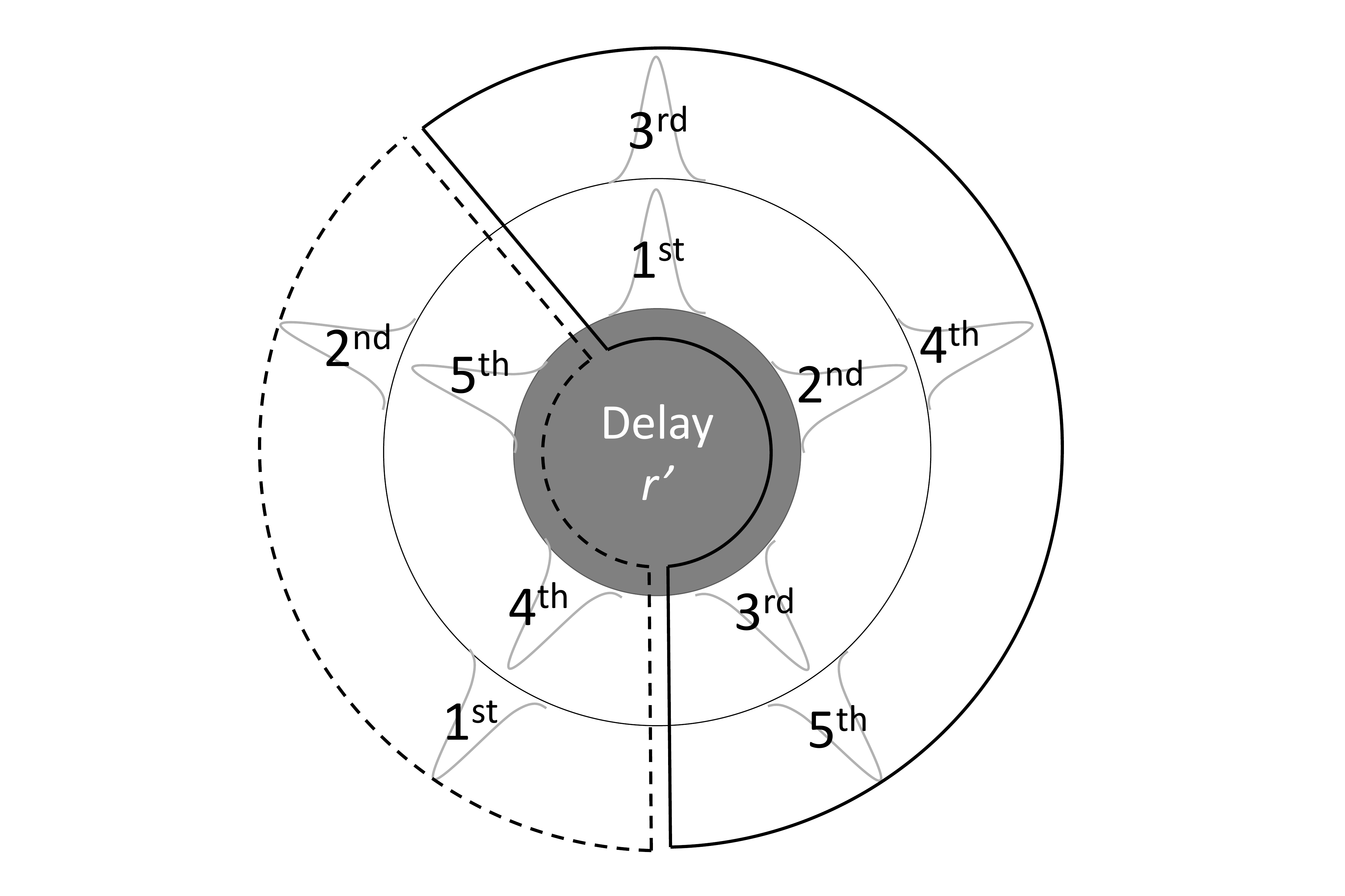}\\
(a) Bob's alternative (dial) measurement
\end{center}
\end{minipage}

\begin{minipage}{0.5\hsize}
\begin{center}
\includegraphics[width=8cm]{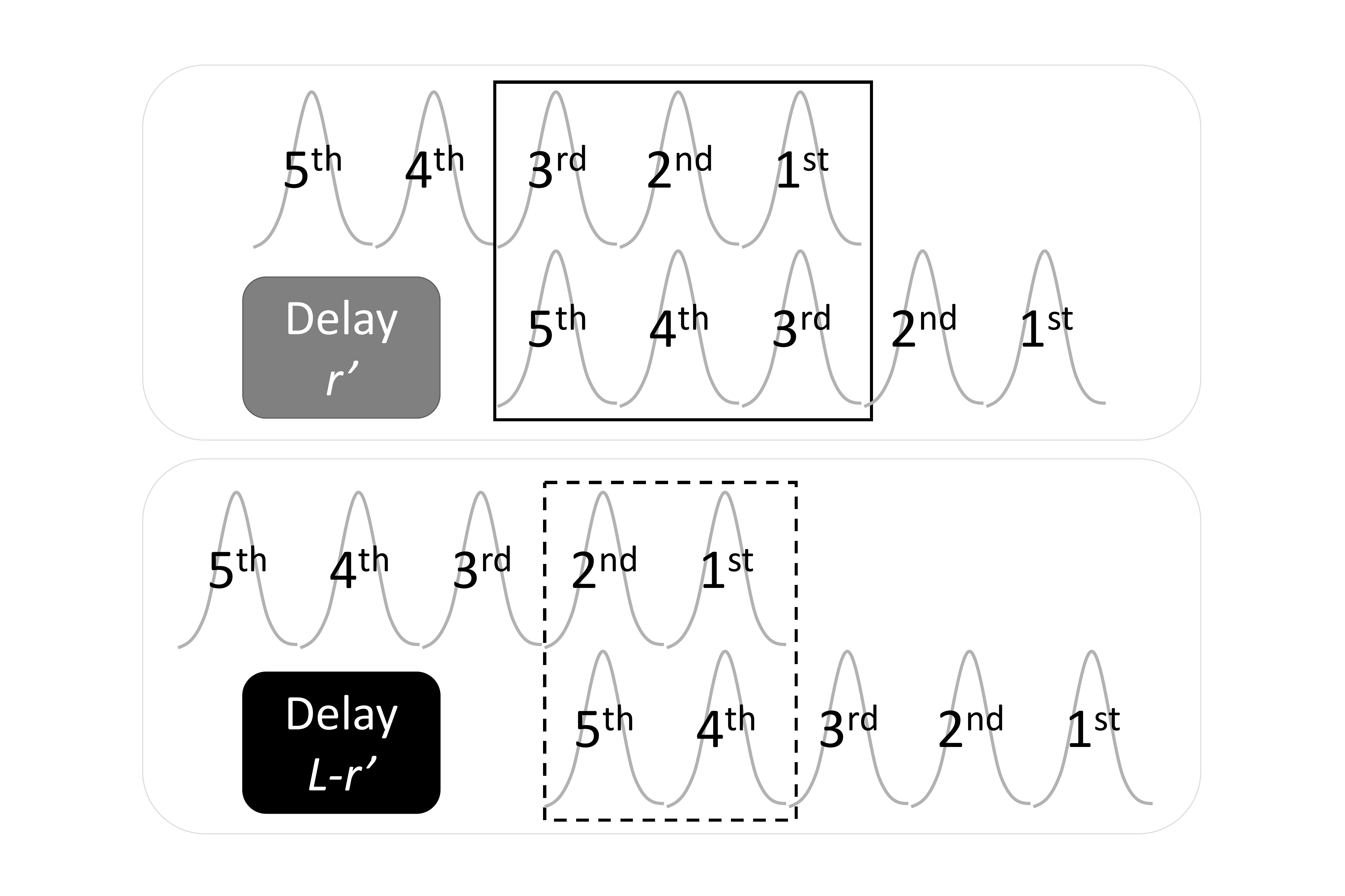}\\
(b) Bob's actual measurement
\end{center}
\end{minipage}

\end{tabular}
\caption{Schematics of (a) Bob's ``dial measurement" with $L=5$ and the delay $r'=2$ and (b) the corresponding actual measurement.
In the dial measurement, there are five patterns of successful detection events. 
Concretely, he obtains the relative phase between one of the following five pairs of interfering pulses: ($1^\mathrm{st}$, $3^\mathrm{rd}$), ($2^\mathrm{nd}$, $4^\mathrm{th}$), ($3^\mathrm{rd}$, $5^\mathrm{th}$) represented by the regime surrounded by the solid line, ($1^\mathrm{st}$, $4^\mathrm{th}$), or ($2^\mathrm{nd}$, $5^\mathrm{th}$) represented by the regime surrounded by the dashed line.
In the actual measurement, by using a specific delay $r'$ or $L-r'$, he can obtain a part of information that he could obtain in his dial measurement with the delay $r'$.
For example, if he employs the delay $r'~(=2)$ in (b), he can obtain the relative phase between either ($1^\mathrm{st}$, $3^\mathrm{rd}$), ($2^\mathrm{nd}$, $4^\mathrm{th}$) or ($3^\mathrm{rd}$, $5^\mathrm{th}$) interfering pulses in (a).
These events correspond to those in the regime surrounded by the solid line in (b).
Also, if he employs the delay $L-r'~(=3)$ in (b), he can obtain the relative phase between either ($1^\mathrm{st}$, $4^\mathrm{th}$) or ($2^\mathrm{nd}$, $5^\mathrm{th}$) interfering pulses in (a).
These events correspond to those in the regime surrounded by the dashed line in (b).
Therefore, if he switches the delay $r'$ or $L-r'$ uniformly at random in his actual measurement, he can simulate the dial measurement with the delay $r'$, while the detection efficiency of his actual measurement is half that of his dial measurement (see Lemma~\ref{fact:dial-actual-relation} below).}
\label{fig:dial_measurement}
\end{center}
\end{figure*}
\subsection{Bob's alternative measurement}
\label{subsec:dial-measurement}
In this subsection, we introduce Bob's alternative measurement, which we will employ in the security proof. 
In step (A2), Bob extracts the \textit{detected} events, in which only one photon is contained in the incoming $L$ pulses. Here, $\{\ket{k}_\textrm{B}\}_{k=1}^L$ denotes the set of basis vectors of the Hilbert space $\mathcal{H}_\textrm{B}$, and $\ket{k}_\textrm{B}$ represents that the $k^\mathrm{th}$ pulse sent is in a single-photon state. Let $\{\hat{B}_{k, s}^{(r)}\}_{k, s}$ be the POVM for the bit value $s$ detected at the pair of ($k^\mathrm{th}$, $(k+r)^\mathrm{th}$)-interfering pulses under the condition that the delay $r$ is chosen.
Considering the effect of the 50:50 BS, $\{\hat{B}_{k, s}^{(r)}\}$ is written as
\begin{align}
\hat{B}_{k, s}^{(r)}:=\frac{1}{2}\hat{P}\left(\frac{\ket{k}_\textrm{B}+(-1)^s\ket{k+r}_\textrm{B}}{\sqrt{2}}\right),\label{defi:Bob-actual-measurement}
\end{align}
for $1\leq k\leq L-r$.
Here, we define $\hat{P}(\ket{\phi})\equiv\ketbra{\phi}{\phi}$. From Eq.~\eqref{defi:Bob-actual-measurement}, the probability of obtaining the bit value $s$ and the pair of interfering pulses $(k, k+r)$ in his measurement with the delay $r$ is given by $\tr{\hat{\rho}\hat{B}_{k, s}^{(r)}}$ for an arbitrary state $\hat{\rho}$ given that exactly one photon is contained in the $L$ pulses.

Next, for simplicity of the security analysis, we convert Bob's actual measurement into the alternative one. We call it the dial measurement, which gives the relative phase of an arbitrary pair of $(i^\textrm{th}, j^\textrm{th})$ $(i<j)$ interfering pulses such that $j-i=r~\textrm{or}~L-r$ for given $r$ (see FIG.~\ref{fig:dial_measurement}(a)).
This measurement has more symmetry than the actual measurement, which makes our analysis much simpler, and importantly it is equivalent to the actual measurement except for $50\%$ of losses as we explain in Lemma~\ref{fact:dial-actual-relation} below.
The POVM $\{\hat{E}_{k, s}^{(r)}\}_{k, s}$ of the dial measurement with the delay $r$ is defined by 
\begin{align}
\hat{E}_{k, s}^{(r)}:=\frac{1}{2}\hat{P}\left(\frac{\ket{k}_\textrm{B}+(-1)^s\ket{k+_Lr}_\textrm{B}}{\sqrt{2}}\right),\label{eq:dial-measurement-povm}
\end{align}
for $1\leq k\leq L$. Here, $+_L$ denotes the summation in modulo $L$, namely, for integers $(p, q)$ with $1\leq p\leq L$ and $1\leq q \leq L$,
\begin{align}
p+_Lq=
\begin{cases}
p+q&\textrm{if }p+q\leq L,\\
p+q-L&\textrm{if }p+q\geq L+1.
\end{cases}
\end{align}
If Bob performs the dial measurement with the delay $r$, the probability of obtaining the bit value $s$ and the pair of interfering pulses $(k, k+_Lr)$ is given by $\tr{\hat{\rho}\hat{E}_{k, s}^{(r)}}$. Note that the following relation holds for $\hat{E}_{k,s}^{(r)}$:
\begin{align}
\hat{E}_{k+_Lr, s}^{(L-r)}=\hat{E}_{k, s}^{(r)}.\label{eq:dial_r_L-r}
\end{align}
Next, we introduce the following lemma~\cite{sasaki2014practical} that relates the dial and actual measurements.
See Appendix~\ref{append:dial-measurement} for its proof. 
\begin{lem}
We define two conditional probabilities $\cpr{s\wedge(i, j)}{r'}_\mathrm{dial}$ and $\cpr{s\wedge(i,j)}{r\in\{r', L-r'\}}_\mathrm{actual}$. $\cpr{s\wedge(i, j)}{r'}_\mathrm{dial}$ represents the probability that Bob obtains the bit value $s$ from $(i^\mathrm{th}, j^{\mathrm{th}})~~(i<j)$ interfering pulses given that he performs the dial measurement with the delay $r=r'\in\mathcal{R}$.
$\cpr{s\wedge(i,j)}{r\in\{r', L-r'\}}_\mathrm{actual}$ represents the probability that Bob obtains $s$ from $(i^\mathrm{th}, j^{\mathrm{th}})$ interfering pulses given that he performs the actual measurement with the delays $r=r'$ or $r=L-r'$ chosen uniformly at random. Then, for an arbitrary fixed $r'\in\mathcal{R}$ and for any input state $\hat{\rho}$,
\begin{align}
\cpr{s\wedge(i, j)}{r'}_\mathrm{dial}=2\cpr{s\wedge(i,j)}{r\in\{r', L-r'\}}_\mathrm{actual}\label{eq:dial-actual}
\end{align}
holds.
\label{fact:dial-actual-relation}
\end{lem}
Lemma~\ref{fact:dial-actual-relation} means that, the dial measurement with the delay $r'$ after performing a half transmittance filter has the same probability distribution of $s$ and $(i,j)$ as the one of the actual measurement when Bob switches delays $r'$ and $L-r'$ uniformly at random (see FIG.~\ref{fig:dial_measurement}(b)). In other words, Eve cannot distinguish which of the measurement was actually employed from the classical information announced by Bob. Thanks to Eq.~\eqref{eq:dial-actual}, we are allowed to use the dial measurement for proving the security of the actual protocol.
We call the protocol where Bob performs the dial measurement instead of the actual one the \textit{alternative protocol}. The alternative protocol runs the same as the actual protocol except for steps (A2) and (A3), which are replaced with the following steps (A2') and (A3'), respectively.
\begin{enumerate}[label=(A\arabic*')]
\setlength{\parskip}{0cm}
\setlength{\itemsep}{0cm}
\setcounter{enumi}{1}
\item Bob receives the incoming $L$ pulses and splits them into two $L$-pulse blocks by using the $50:50$ BS.
He selects a delay $r$ uniformly at random from the set $\mathcal{R}\subset\{1, 2, \dots, L-1\}$. After that, Bob performs the dial measurement. Let us call the event \textit{detected} if he detects exactly one photon in the pair of interfering pulses $(k^\textrm{th}, (k+_Lr)^\textrm{th})$ ($1\leq k\leq L$), and detects the vacuum in all the other pairs of interfering pulses. If the event is not \textit{detected}, Alice and Bob skip steps (A3') and (A4).
\item Bob takes note of the detected bit value and announces the pair of numbers $(i, j)=(\min\{k, k+_Lr\}, \max\{k, k+_Lr\})$ over an authenticated public channel.
\end{enumerate}

\subsection{Virtual protocol}
\label{subsec:virtual-protocol}
In this subsection, we introduce the entanglement distillation protocol to prove the security of the alternative protocol. Our analysis is based on the Shor-Preskill's security proof~\cite{shor2000simple}, where we follow similar arguments of the security proof of the original DPS protocol~\cite{tamaki2012unconditional}.
To show Alice and Bob virtually extract a maximally entangled state, we need to introduce ancilla systems on Alice's side and decompose Bob's measurement, which we explain below. 

\begin{figure}[t]
\begin{center}
\includegraphics[width=8cm]{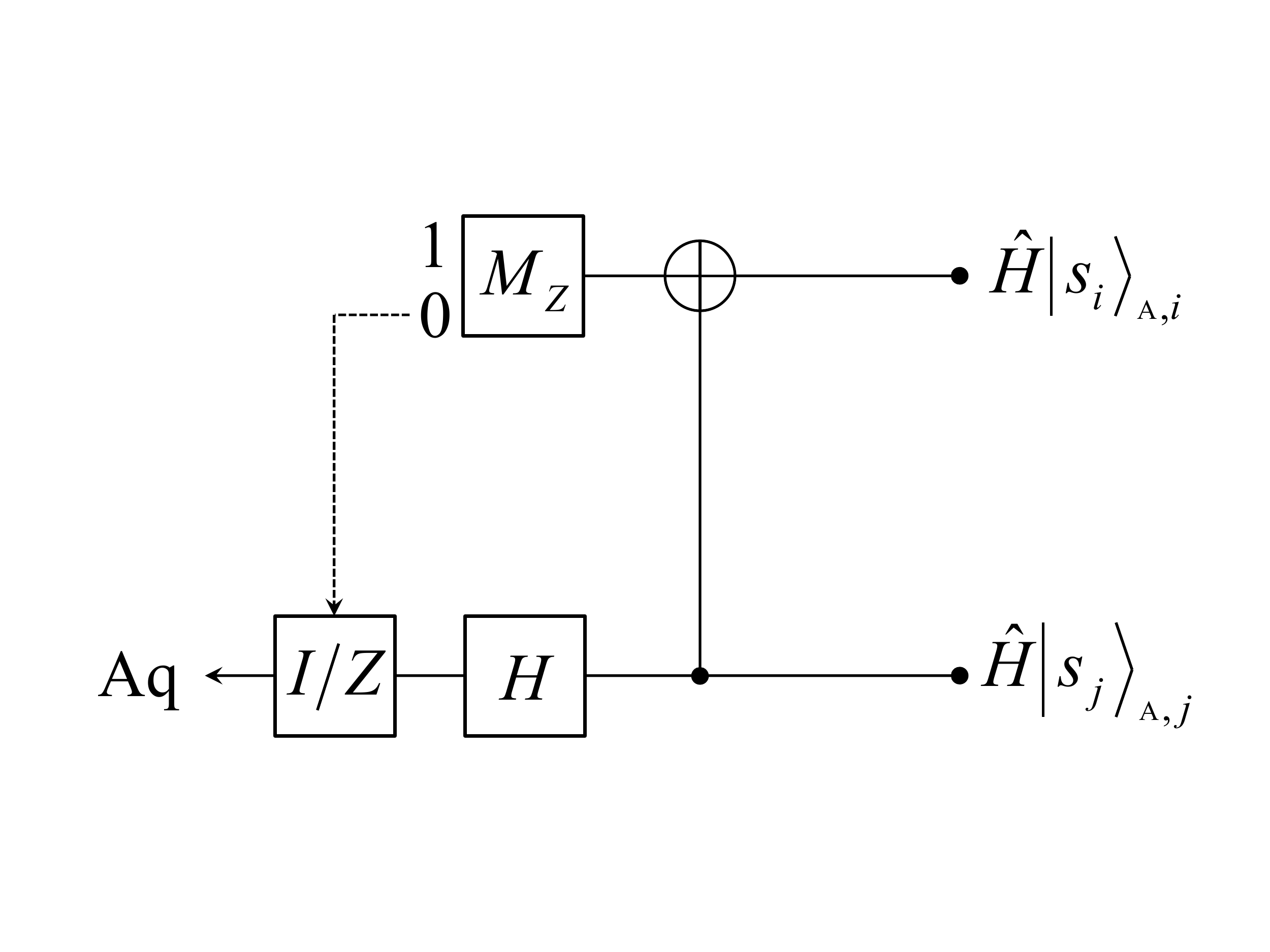}
\caption{The quantum circuit representing Alice's procedure in the virtual protocol. The inputs are the $i^\mathrm{th}$ and $j^\mathrm{th}$ qubits, where the pair of integers $(i, j)~(i<j)$ is announced by Bob. She applies a C-NOT gate (defined on the Z basis $\{\ket{0}_\textrm{A},\ket{1}_\textrm{A}\}$) to these qubits, and the $j^\mathrm{th}$ qubit is subjected to a Hadamard gate (denoted by $H$) while the $i^\mathrm{th}$ qubit is measured on $Z$-basis (denoted by $M_Z$). If the outcome of the $Z$ measurement is $0$, she applies a phase flip gate to the $j^\mathrm{th}$ qubit with probability $1/2$ ,which is denoted by $I/Z$. Otherwise, she applies the identity operation to the $j^\mathrm{th}$ qubit. After that, we name the quantum system of the $j^\mathrm{th}$ qubit Aq.}
\label{fig:quantum_circuit}
\end{center}
\end{figure}

First, we explain Alice's sending state in the virtual protocol.
Suppose Alice has a quantum register of $L$-qubit system and let $\mathcal{H}_\textrm{A}=\bigotimes_{k=1}^{L}\mathcal{H}_{\textrm{A}, k}$ be the Hilbert space of these systems. Then, Alice's state preparation is equivalent to the preparation of the following state over the quantum register system and $L$ pulses as 
\begin{align}
\ket{\Phi_\delta}:=2^{-L/2}\sum_{\vec{s}}\bigotimes_{k=1}^L\left(\hat{H}\ket{s_k}_{\textrm{A},k}\right)\ket{(-1)^{s_k}\mathrm{e}^{\mathrm{i}\delta}\alpha}_k,\label{defi:alice_prepare_state_entangle}
\end{align}
where $\hat{H}\equiv\frac{1}{\sqrt{2}}\sum_{s, s'=0, 1}(-1)^{ss'}\ketbra{s}{s'} $ denotes the Hadamard operator. Note that $\delta\in[0, 2\pi)$ is chosen uniformly at random for each preparation of the state $\ket{\Phi_\delta}$.
As shown in step (A4), the information that Alice needs to obtain is $s_i\oplus s_j$. 
To obtain this information, she applies the following quantum circuit (see FIG.~\ref{fig:quantum_circuit}) to the qubits $i$ and $j$ upon receiving from Bob, and measures the qubit Aq in the computational basis $\{\ket{0}_\textrm{Aq}, \ket{1}_\textrm{Aq}\}$. The set of measurement operators that Alice performs can be represented by
\begin{align}
\hat{M}_1^{(i, j)}=&\hat{H}\ket{0}_\textrm{Aq}\left({}_{\textrm{A}, i}\bra{0}{}_{\textrm{A}, j}\bra{1}\right)+\hat{H}\ket{1}_\textrm{Aq}\left({}_{\textrm{A}, i}\bra{1}{}_{\textrm{A}, j}\bra{0}\right),\nonumber\\
\hat{M}_2^{(i, j)}=&\frac{1}{\sqrt{2}}\ket{0}_\textrm{Aq}\left({}_{\textrm{A}, i}\bra{0}{}_{\textrm{A}, j}\bra{0}+{}_{\textrm{A}, i}\bra{1}{}_{\textrm{A}, j}\bra{1}\right),\nonumber\\
\hat{M}_3^{(i, j)}=&\frac{1}{\sqrt{2}}\ket{1}_\textrm{Aq}\left({}_{\textrm{A}, i}\bra{0}{}_{\textrm{A}, j}\bra{0}-{}_{\textrm{A}, i}\bra{1}{}_{\textrm{A}, j}\bra{1}\right).
\end{align}
Note that Alice's state preparation of $\ket{\Phi_\delta}$ with a random and uniform $\delta$ followed by the measurement $\{M_{m}^{(i, j)}\}_{m=1,2,3}$ is equivalent to the step (A1).
Moreover, in the virtual protocol, instead of $\ket{\Phi_\delta}$, Alice prepares the following state for simplicity of analysis.
\begin{align}
\ket{\Phi}:=2^{-L/2}\sum_{\vec{s}}\sum_{\nu=0}^\infty\ket{\nu}_\textrm{C}\hat{\pi}_\nu\bigotimes_{k=1}^L\left(\hat{H}\ket{s_k}_{\textrm{A},k}\right)\ket{(-1)^{s_k}\alpha}_k.\label{defi:alice_prepare_state_phase_randomized}
\end{align}
Here, C is the system that stores the number of photons contained in the $L$ pulses whose Hilbert space is spanned by an orthogonal basis $\{\ket{\nu}_\textrm{C}\}_{\nu=0}^\infty$. Also, $\hat{\pi}_\nu$ is the projection onto the subspace that the total photon number in the $L$ pulses is $\nu$. 
From Eve's perspective, accessible quantum information of Eqs.~\eqref{defi:alice_prepare_state_entangle} and ~\eqref{defi:alice_prepare_state_phase_randomized} are the same since the following equation holds. 
\begin{align}
\frac{1}{2\pi}\int_0^{2\pi}\mathrm{d}\delta\ketbra{\Phi_\delta}{\Phi_\delta}=\mathrm{Tr}_\textrm{C}\ketbra{\Phi}{\Phi}.
\end{align}

Next, we explain Bob's measurement procedure in the virtual protocol.
In principle, he is able to determine whether the event is \textit{detected} or not before he determines the pair of interfering pulses and the bit value by performing the quantum nondemolition (QND) measurement of the total photon number in the incoming $L$ pulses. The event is called \textit{detected} if and only if the measurement outcome of the QND measurement is exactly one photon in the block of $L$ pulses. In the \textit{detected} events, the dial measurement is decomposed into two measurements, namely, the POVM in Eq.~\eqref{eq:dial-measurement-povm} is decomposed into
\begin{align}
\hat{E}_{k, s}^{(r)}=\hat{F}_{k}^{(r)\dagger} \hat{P}(\ket{s}_\textrm{Bq})\hat{F}_{k}^{(r)}.
\label{eq:bob_decoposition}
\end{align}
Here, the set of measurement operators $\hat{F}_{k}^{(r)}: \mathcal{H}_\textrm{B}\to\mathcal{H}_\textrm{Bq}$ represents a filtering operation that gives the outcome $k$ and leaves a qubit system $\mathcal{H}_\textrm{Bq}$, which is defined by
\begin{align}
\hat{F}_{k}^{(r)}:=\left\{
\begin{array}{l}
\displaystyle\frac{1}{\sqrt{2}}\hat{H}\ket{1}_\textrm{Bq}{}_\textrm{B}\bra{k}+\frac{1}{\sqrt{2}}\hat{H}\ket{0}_\textrm{Bq}{}_\textrm{B}\bra{k+r}\\
\hfill\textrm{if }1\leq k\leq L-r,\\
\displaystyle\frac{1}{\sqrt{2}}\hat{H}\ket{1}_\textrm{Bq}{}_\textrm{B}\bra{k+r-L}+\frac{1}{\sqrt{2}}\hat{H}\ket{0}_\textrm{Bq}{}_\textrm{B}\bra{k}\\
\hfill\textrm{if }L-r+1\leq k\leq L.
\end{array}
\right.
\label{defi:filtering-operation}
\end{align}

By using Alice's quantum circuit represented by FIG.~3 and Bob's filtering operation described by $\{\hat{F}_k^{(r)}\}_k$, we introduce the following entanglement distillation protocol (EDP).

\begin{enumerate}[label=(V\arabic*)]
\setlength{\parskip}{0cm}
\setlength{\itemsep}{0cm}
\item Alice prepares $\ket{\Phi}$ and sends a block of $L$ pulses to Bob through a quantum channel.
\item Bob receives the incoming $L$ pulses and performs the QND measurement of the total photon number in the $L$ pulses. Let us call the event \textit{detected} if he detects exactly one photon in the block of $L$ pulses. If the event is not \textit{detected}, Alice and Bob skip steps (V3) and (V4) below.
\item Bob chooses $r$ uniformly at random from the set $\mathcal{R}=\bigcup_{m=1}^t\{m, L-m\}$, and performs the filtering operation $\{\hat{F}^{(r)}_k\}$. He obtains the pair of pulses $(k, k+_Lr)$ where exactly one photon is contained, and obtains the output qubit Bq.
He sends the pair of integers $(i, j):=(\min\{k, k+_Lr\}, \max\{k, k+_Lr\})$ to Alice over an authenticated public channel.
\item Alice applies the quantum circuit in FIG.~\ref{fig:quantum_circuit} on her $i^\mathrm{th}$ and $j^\mathrm{th}$ qubits, and outputs the qubit Aq. Also, she measures system C and learns the total photon number $\nu$ in the block of $L$ pulses.
\item Alice and Bob repeat steps (V1) through (V4) for $N$ times. Let $NQ$ be the number of the \textit{detected} events. At this point, Alice and Bob share $NQ$ pairs of qubits.
\item Alice and Bob randomly select a small portion $\xi$ of the $NQ$ \textit{detected} events, measure the qubits in the computational basis $\{\ket{0}, \ket{1}\}$, and compare the bit values over the public channel. This gives the estimate of the bit error rate $e^{(\textrm{b})}$ and hence of the number $e^{(\textrm{b})}NQ(1-\xi)$ of bit errors in the remaining portion.
\item Alice and Bob discuss over the public channel to perform entanglement distillation on the remaining pairs of qubits. Finally, they measure all the remaining pairs of qubits on the computational basis to obtain a final key of length $GN(1-\xi)$.
\end{enumerate}

In this protocol, the key generation rate per sending pulse is written as~\cite{gottesman2004security}
\begin{align}
G=[Q(1-h(e^{(\textrm{b})}))-h^{(\textrm{ph})}]/L,\label{eq:keyrate}
\end{align}
where $h(x)=-x\log_2x-(1-x)\log_2(1-x)$, and $h^{(\textrm{ph})}$ expresses the number of privacy amplification. The explicit formula of $h^{(\textrm{ph})}$ is described by
\begin{align}
h^{(\textrm{ph})}=\sum_{\nu=0}^\infty Q^{(\nu)}h(e^{(\textrm{ph}, \nu)}),\label{eq:privacy-amplification}
\end{align}
where $Q^{(\nu)}$ denotes the function of \textit{detected} events when Alice emits $\nu$ photons satisfying $\sum_{\nu=0}^\infty Q^{(\nu)}=Q$, and $e^{(\textrm{ph}, \nu)}$ denotes the phase error rate when Alice emits $\nu$ photons. Note that the equivalence between steps (A7) and (V7) is guaranteed by the discussion in~\cite{shor2000simple}. Since the phase error rate $e^{(\textrm{ph}, \nu)}$ cannot be obtained directly in the actual protocol, we need to estimate $e^{(\textrm{ph}, \nu)}$ with some statistics such as the disturbance information during Alice and Bob's quantum communication.

\subsection{POVM elements for the bit and phase errors}
\label{subsec:POVM-elements}
In this subsection, we construct POVMs for the bit and phase errors to estimate the upper bound on the phase error rate. To derive the relation between the bit and the phase error rates, we consider a measurement on Alice and Bob's quantum registers A and B just after the event is \textit{detected} at step (V2), and regard an outcome as the occurrence of a bit error or a phase error. In this subsection, we explain only the definition and the resulting forms of POVMs for the bit and phase errors.
The detailed derivations are referred to Appendix~\ref{append:bit-phase-error-POVMs}.

The POVM element corresponding to the bit error in the pair of pulses $(i, j)$ for $i<j$ is defined as
\begin{align}
\begin{split}
\hat{e}^{(\textrm{b})}_{i, j}:=&\sum_{s_i, s_j\in\{0, 1\}}\hat{P}\left(\hat{H}\ket{s_i}_{\textrm{A}, i}\right)\hat{P}\left(\hat{H}\ket{s_j}_{\textrm{A}, j}\right)\\
&\otimes\frac{1}{\abs{\mathcal{R}}}\sum_{r\in\mathcal{R}}\left[\hat{E}_{i, s_i\oplus s_j\oplus1}^{(r)}\delta_{j, i+r}+\hat{E}_{j, s_i\oplus s_j\oplus1}^{(r)}\delta_{i, j+r-L}\right]
\end{split}\nonumber\\
=&\frac{2}{\abs{\mathcal{R}}}\sum_{s_i, s_j\in\{0, 1\}}\hat{P}\left(\hat{H}\ket{s_i}_{\textrm{A}, i}\right)\hat{P}\left(\hat{H}\ket{s_j}_{\textrm{A}, j}\right)\otimes\hat{E}_{i, s_i\oplus s_j\oplus1}^{(j-i)}.
\label{defi:ebitPOVM_ij}
\end{align}
Here, $\abs{\mathcal{R}}$ denotes the cardinality of the set $\mathcal{R}$, and we use Eq.~\eqref{eq:dial_r_L-r} in the last equality.
Note that we omit identity operators on the subsystems.
Next, the POVM element for a phase error is defined by the instances where Alice and Bob measure their qubits with the Hadamard basis $\{\hat{H}\ket{0}, \hat{H}\ket{1}\}$ and their outcomes disagree.
The POVM element corresponding to the occurrence of the phase error in the pair of pulses $(i, j)$ for $i<j$ is given by
\begin{align}
\begin{split}
\hat{e}_{i, j}^{(\textrm{ph})}:=&\sum_{s=0}^1\sum_{k=1}^3\hat{M}_k^{(i, j)\dagger}\hat{P}\left(\hat{H}\ket{s}_\textrm{Aq}\right)\hat{M}_k^{(i, j)}\\
&\otimes\frac{1}{\abs{\mathcal{R}}}\sum_{r\in\mathcal{R}}\left[\hat{F}_i^{(r)\dagger}\hat{P}\left(\hat{H}\ket{\bar{s}}_\textrm{B}\right)\hat{F}_i^{(r)}\delta_{j, i+r}\right.\\
&\left.\quad+\hat{F}_j^{(r)\dagger}\hat{P}\left(\hat{H}\ket{\bar{s}}_\textrm{B}\right)\hat{F}_j^{(r)}\delta_{i, j+r-L}\right]
\end{split}\nonumber\\
\begin{split}
=&\frac{1}{\abs{\mathcal{R}}}\left[\hat{P}\left(\ket{0}_{\textrm{A}, i}\ket{1}_{\textrm{A}, j}\right)\otimes\hat{P}\left(\ket{i}_\textrm{B}\right)\right.\\
&\left.+\hat{P}\left(\ket{1}_{\textrm{A}, i}\ket{0}_{\textrm{A}, j}\right)\otimes\hat{P}\left(\ket{j}_\textrm{B}\right)\right]
\end{split}\nonumber\\
&+\frac{1}{2\abs{\mathcal{R}}}\sum_{t=0}^1\hat{P}\left(\ket{t}_{\textrm{A}, i}\ket{t}_{\textrm{A}, j}\right)\otimes\left[\hat{P}\left(\ket{i}_\textrm{B}\right)+\hat{P}\left(\ket{j}_\textrm{B}\right)\right],
\label{defi:ephPOVM_ij}
\end{align}
where $\bar{s}=s\oplus1$.
For simplicity of analysis, we introduce a unitary operator $\hat{U}$ acting on $\mathcal{H}_\textrm{A}\otimes\mathcal{H}_\textrm{B}$ defined by
\begin{align}
\hat{U}\bigotimes_{k'=1}^L\left(\hat{H}\ket{s_{k'}}_{\textrm{A}, k'}\right)\ket{k}_\textrm{B}=(-1)^{s_k}\bigotimes_{k'=1}^L\left(\hat{H}\ket{s_{k'}}_{\textrm{A}, k'}\right)\ket{k}_\textrm{B}.\label{defi:unitary}
\end{align}
By using $\hat{U}$ and Eq.~\eqref{defi:ebitPOVM_ij}, it is straightforward to show that
\begin{align}
\hat{U}\hat{e}^{(\textrm{b})}_{i, j}\hat{U}^\dagger=\frac{2}{\abs{\mathcal{R}}}\hat{1}_\textrm{A}\otimes\hat{E}_{i, 1}^{(j-i)}=\hat{1}_\textrm{A}\otimes\frac{1}{2\abs{\mathcal{R}}}\hat{P}\left(\ket{i}_\textrm{B}-\ket{j}_\textrm{B}\right).\label{eq:unitary-ebit_ij}
\end{align}
Since $\hat{U}$ also satisfies
\begin{align}
\hat{U}\hat{P}(\ket{s}_{\textrm{A}, k})\hat{P}(\ket{k'}_\textrm{B})\hat{U}^\dagger=\hat{P}(\ket{s\oplus\delta_{k, k'}}_{\textrm{A}, k})\hat{P}(\ket{k'}_\textrm{B}),
\end{align}
we have
\begin{align}
\hat{U}\hat{e}_{i, j}^{(\textrm{ph})}\hat{U}^\dagger=&\frac{1}{2\abs{\mathcal{R}}}\left[\hat{P}(\ket{1}_{\textrm{A}, i})+\hat{P}(\ket{1}_{\textrm{A}, j})\right]\nonumber\\
&\otimes\left[\hat{P}(\ket{i}_\textrm{B})+\hat{P}(\ket{j}_\textrm{B})\right].\label{eq:unitary-eph_ij}
\end{align}

For the state $\hat{\rho}$ of a \textit{detected} event, the probability of having a bit error in the extracted qubit pair $\textrm{Aq}$ and $\textrm{Bq}$ is expressed by $\tr{\hat{\rho}\hat{e}^{(\textrm{b})}}$, while a phase error is given by $\tr{\hat{\rho}\hat{e}^{(\textrm{ph})}}$, where $\hat{e}^{(\textrm{b})}$ and $\hat{e}^{(\textrm{ph})}$ are respectively given by 
\begin{align}
\hat{e}^{(\textrm{b})}=\sum_{(i, j):j-i\in\mathcal{R}}\hat{e}^{(\textrm{b})}_{i, j},~~~~~\hat{e}^{(\textrm{ph})}=\sum_{(i, j):j-i\in\mathcal{R}}\hat{e}_{i, j}^{(\textrm{ph})}.\label{eq:ebit-eph-povm}
\end{align}
By applying $\hat{U}$ to $\hat{e}^{(\textrm{b})}$ in Eq.~\eqref{eq:ebit-eph-povm} and using Eq.~\eqref{eq:unitary-ebit_ij}, we have
\begin{align}
\hat{U}\hat{e}^{(\textrm{b})}\hat{U}^\dagger=\hat{1}_\textrm{A}\otimes\hat{\Pi}^{(\textrm{b})},\label{eq:unitary-ebit}
\end{align}
where the matrix elements of $\hat{\Pi}^{(\textrm{b})}$ are
\begin{align}
{}_\textrm{B}\bra{m}\hat{\Pi}^{(\textrm{b})}\ket{n}_\textrm{B}=
\begin{cases}
\frac{1}{2}&\textrm{if } m=n,\\
-\frac{1}{2\abs{\mathcal{R}}}&\textrm{if } \abs{m-n}\in\mathcal{R}\wedge\abs{m-n}\neq\frac{L}{2},\\
-\frac{1}{\abs{\mathcal{R}}}&\textrm{if } \abs{m-n}\in\mathcal{R}\wedge\abs{m-n}=\frac{L}{2},\\
0&\textrm{otherwise}.
\end{cases}\label{eq:Pi}
\end{align}
By applying $\hat{U}$ to $\hat{e}^{(\textrm{ph})}$, using Eq.~\eqref{eq:unitary-eph_ij} and $\hat{P}\left(\ket{k}_{\textrm{A}, m}\right)=\sum_{\vec{a}}\hat{P}\left(\ket{\vec{a}}_\textrm{A}\right)\delta_{a_m, k}$ with $\vec{a}=a_1a_2\dots a_L$ and $\ket{\vec{a}}_\textrm{A}=\ket{a_1}_{\textrm{A}, 1}\ket{a_2}_{\textrm{A}, 2}\dots\ket{a_L}_{\textrm{A}, L}$, $\hat{U}\hat{e}^{(\textrm{ph})}\hat{U}^\dagger$ results in the following form.
\begin{align}
&\hat{U}\hat{e}^{(\textrm{ph})}\hat{U}^\dagger\nonumber\\
\begin{split}
&=\sum_{\vec{a}}\hat{P}\left(\ket{\vec{a}}_\textrm{A}\right)\\
&\quad\otimes\sum_{m=1}^L\hat{P}\left(\ket{m}_\textrm{B}\right)\left(\frac{1}{2}\delta_{a_m, 1}+\frac{1}{2\abs{\mathcal{R}}}\sum_{n:\abs{m-n}\in\mathcal{R}}\delta_{a_n, 1}\right)
\end{split}\nonumber\\
&=:\sum_{\vec{a}}\hat{P}\left(\ket{\vec{a}}_\textrm{A}\right)\otimes\hat{\Pi}_{\vec{a}}^{(\textrm{ph})}.\label{eq:unitary-eph}
\end{align}

\subsection{Relations between the bit and the phase error rates}
\label{subsec:relation-bit-phase-errors}
In this subsection, we derive the upper bound on $h^{(\textrm{ph})}$ in Eq.~\eqref{eq:keyrate} by using the bit error rate.
For this, we first derive the range where Alice's sending state can be contained.
In the virtual protocol, if the initial state $\ket{\Phi}$ satisfies $\bra{\Phi}\ket{\vec{a}}_\textrm{A}\ket{\nu}_\textrm{C}=0$ for a state $\ket{\vec{a}}_\textrm{A}\ket{\nu}_\textrm{C}$, the density operator $\hat{\rho}$ of Aq and Bq originating from $\ket{\Phi}$ also satisfies $\hat{\rho}\ket{\vec{a}}_\textrm{A}\ket{\nu}_\textrm{C}=0$.
Moreover, we have the following relations between $\vec{a}$ and $\nu$ such that $\bra{\Phi}\ket{\vec{a}}_\textrm{A}\ket{\nu}_\textrm{C}=0$ is satisfied~\cite{tamaki2012unconditional},
\begin{align}
{}_\textrm{A}\bra{\vec{a}}_\textrm{C}\bra{\nu}\ket{\Phi}&=0~~~\textrm{if }\abs{\vec{a}}>\nu,\label{eq:constraint-single-photon}\\
{}_\textrm{A}\bra{\vec{a}}_\textrm{C}\bra{\nu}\ket{\Phi}&=0~~~\textrm{if }(-1)^{\abs{\vec{a}}}\neq(-1)^\nu,\label{eq:constraint-photon-parity}
\end{align}
where $\abs{\vec{a}}$ denotes the number of 1's in the bit string $\vec{a}$.
By using Eqs.~\eqref{eq:constraint-single-photon} and~\eqref{eq:constraint-photon-parity}, $\hat{\rho}$ after Alice obtains $\nu$ by measuring the system C is contained in the range of a projection operator $\hat{P}^{(\nu)}$, which is defined by
\begin{align}
\hat{P}^{(\nu)}:=\sum_{\vec{a}: \abs{\vec{a}}=\nu, \nu-2, \nu-4, \dots}\sum_{i=1}^L\hat{P}\left(\ket{\vec{a}}_\textrm{A}\otimes\ket{i}_\textrm{B}\right).
\end{align}

Next, to derive the relation between the bit and phase error rates, we consider the quantity $\Omega^{(\nu)}(\lambda)$ defined as the largest eigenvalue of the operator
\begin{align}
\hat{P}^{(\nu)}\left(\hat{e}^{(\textrm{ph})}-\lambda\hat{e}^{(\textrm{b})}\right)\hat{P}^{(\nu)}\label{eq:relation-ebit-eph-POVM}
\end{align}
in the range of $\hat{P}^{(\nu)}$ with $\lambda\geq0$. By using $\Omega^{(\nu)}(\lambda)$, the phase error rate $e^{(\textrm{ph}, \nu)}$ when Alice emits $\nu$ photons is bounded by the bit error rate $e^{(\textrm{b}, \nu)}$ when Alice emits $\nu$ photons as~\cite{tamaki2012unconditional}
\begin{align}
e^{(\textrm{ph}, \nu)}\leq\lambda e^{(\textrm{b}, \nu)}+\Omega^{(\nu)}(\lambda).\label{eq:eph-nu-bound}
\end{align}
Since Eq.~\eqref{eq:eph-nu-bound} for various $\lambda\geq0$ determines a convex achievable region of $(e^{(\textrm{b}, \nu)}, e^{(\textrm{ph}, \nu)})$ and $h(x)$ is monotonically increasing, we obtain the convex achievable region of $(e^{(\textrm{b}, \nu)}, h(e^{(\textrm{ph}, \nu)}))$ specified by
\begin{align}
h(e^{(\textrm{ph}, \nu)})\leq\gamma e^{(\textrm{b}, \nu)}+\Omega_h^{(\nu)}(\gamma)\label{eq:heph-nu-bound}
\end{align}
for various $\gamma\geq0$. Here, $\Omega_h^{(\nu)}(\gamma)$ is the quantity depending on $\gamma$ and $\Omega^{(\nu)}(\lambda)$.

In order to derive an upper bound on the leaked information $h^{(\textrm{ph})}$ in Eq.~\eqref{eq:privacy-amplification}, we consider the optimization of $Q^{(\nu)}$ such that Eve's information is maximal. Since $NQ^{(\nu)}$ is the number of qubits extracted in the step (V5) from the \textit{detected} events when Alice emits $\nu$ photons, $Q^{(\nu)}$ needs to satisfy the following physical requirement regarding the number of total events when Alice emits $\nu$ photons
\begin{align}
NQ^{(\nu)}\leq Np_\nu,\label{ineq:Q-const}
\end{align}
where $p_\nu$ denotes the Poisson distribution with mean $L\alpha^2$,
\begin{align}
p_\nu:=e^{-L\alpha^2}\frac{(L\alpha^2)^\nu}{\nu!}.
\end{align}
Let $q^{(\nu)}$ be a fraction of \textit{detected} events when Alice emits $\nu$ photons 
among all the detection,
\begin{align}
q^{(\nu)}=\frac{Q^{(\nu)}}{Q}=\frac{Q^{(\nu)}}{\sum_{\nu=0}^\infty Q^{(\nu)}}.
\end{align}
Here, $q^{(\nu)}$ is chosen by Eve under the constraint of Eq.~\eqref{ineq:Q-const}.
As long as $\Omega^{(0)}(\lambda)\leq\Omega^{(1)}(\lambda)\leq\Omega^{(2)}(\lambda)\leq\dots$ holds for all $\lambda\geq0$, Eve can maximize the amount of leaked information by using the events with a larger value of $\nu$.
Therefore, the optimal strategy for Eve is the following choice:
\begin{align}
q^{(\nu)}=
\begin{cases}
Q^{-1}p_\nu&\textrm{if }\nu\geq \nu_0+1\\
1-Q^{-1}(1-\sum_{\nu'=0}^{\nu_0}p_{\nu'})&\textrm{if }\nu=\nu_0\\
0&\textrm{if }\nu\leq\nu_0-1,
\end{cases}
\end{align}
where $\nu_0$ is the integer satisfying
\begin{align}
1-\sum_{\nu'=0}^{\nu_0}p_{\nu'}<Q\leq1-\sum_{\nu'=0}^{\nu_0-1}p_{\nu'}.
\end{align}
By using $\{q^{(\nu)}\}_\nu$ and Eq.~\eqref{eq:heph-nu-bound}, the upper bound on $h^{(\textrm{ph})}$ in Eq.~\eqref{eq:privacy-amplification} is written as
\begin{align}
h^{(\textrm{ph})}=&\sum_{\nu=0}^\infty q^{(\nu)}h(e^{(\textrm{ph}, \nu)})\nonumber\\
\leq&\min_{\gamma\geq0}\left\{\gamma e^{(\textrm{b})}+\sum_{\nu=0}^\infty q^{(\nu)}\Omega_h^{(\nu)}(\gamma)\right\},\label{eq:heph-bound-tight}
\end{align}
where $e^{(\textrm{b})}=\sum_{\nu}q^{(\nu)}e^{(\textrm{b}, \nu)}$ is the bit error rate in the actual protocol. 
The task left to obtain the upper bound on $h^{(\textrm{ph})}$ is to evaluate the quantities $\Omega^{(\nu)}(\lambda)$ for $\nu\in[0,\infty)$.
In our analysis, we consider the upper bounds on $\Omega^{(\nu)}(\lambda)$ for $\nu=1,2$, and for $\nu\geq3$ we make a pessimistic assumption that all the information is leaked to Eve, that is, $\Omega^{(\nu)}(\lambda)\equiv1$.
With this consideration, Eq.~\eqref{eq:heph-bound-tight} is upper bounded by 
\begin{align}
h^{(\textrm{ph})}\leq\min_{\gamma\geq0}\left\{\gamma e^{(\textrm{b})}+\sum_{\nu=0}^{2}q^{(\nu)}\Omega_h^{(\nu)}(\gamma)+q^{(\nu\geq3)}\right\}
\end{align}
with $q^{(\nu\geq3)}:=\sum_{\nu=3}^{\infty}q^{(\nu)}$.

\subsection{Evaluation of $\Omega^{(\nu)}(\lambda)$}
\label{subsec:evaluation-omega}
To evaluate $\Omega^{(\nu)}(\lambda)$, we apply the unitary operation $\hat{U}$ in Eq.~\eqref{defi:unitary} to Eq.~\eqref{eq:relation-ebit-eph-POVM}, and we obtain~\cite{tamaki2012unconditional}
\begin{align}
&\hat{U}\hat{P}^{(\nu)}\hat{U}^\dagger\left(\hat{U}\hat{e}^{(\textrm{ph})}\hat{U}^\dagger-\lambda\hat{U}\hat{e}^{(\textrm{b})}\hat{U}^\dagger\right)\hat{U}\hat{P}^{(\nu)}\hat{U}^\dagger\nonumber\\
=&\sum_{\vec{a}:\abs{\vec{a}}=\nu-1, \nu-3, \dots}\hat{P}\left(\ket{\vec{a}}_\textrm{A}\right)\otimes(\hat{\Pi}_{\vec{a}}^{(\textrm{ph})}-\lambda\hat{\Pi}^{(\textrm{b})})\nonumber\\
&+\sum_{\vec{a}:\abs{\vec{a}}=\nu+1}\hat{P}\left(\ket{\vec{a}}_\textrm{A}\right)\otimes\hat{P}_{\vec{a}}(\hat{\Pi}_{\vec{a}}^{(\textrm{ph})}-\lambda\hat{\Pi}^{(\textrm{b})})\hat{P}_{\vec{a}}.
\end{align}
Here, we use the following equation
\begin{align}
\hat{U}\hat{P}^{(\nu)}\hat{U}^\dagger=&\sum_{\vec{a}: \abs{\vec{a}}=\nu-1, \nu-3, \dots}\hat{P}\left(\ket{\vec{a}}_\textrm{A}\right)\otimes\hat{1}_\textrm{B}\nonumber\\
&+\sum_{\vec{a}: \abs{\vec{a}}=\nu+1}\hat{P}\left(\ket{\vec{a}}_\textrm{A}\right)\otimes\hat{P}_{\vec{a}},
\end{align}
where
\begin{align}
\hat{P}_{\vec{a}}:=\sum_{i=1}^L\delta_{a_i, 1}\hat{P}\left(\ket{i}_\textrm{B}\right).
\end{align}
Since $\hat{\Pi}_{\vec{a}}^{(\textrm{ph})}\geq\hat{\Pi}_{\vec{a}'}^{(\textrm{ph})}$ for $(\vec{a}, \vec{a}')$ such that $a_i\geq a'_i$ for all $i$, we can neglect the operators with $\vec{a}$ satisfying $\abs{\vec{a}}\leq\nu-3$.
Therefore, we have only to consider $\Omega_{-}^{(\nu)}(\lambda)$ defined as the largest eigenvalue of the operators
\begin{align}
\{\hat{\Pi}_{\vec{a}}^{(\textrm{ph})}-\lambda\hat{\Pi}^{(\textrm{b})}\mid\abs{\vec{a}}=\nu-1\},\label{set:omegaminus}
\end{align}
and $\Omega_{+}^{(\nu)}(\lambda)$ defined as the largest eigenvalue of the operators
\begin{align}
\{\hat{P}_{\vec{a}}(\hat{\Pi}_{\vec{a}}^{(\textrm{ph})}-\lambda\hat{\Pi}^{(\textrm{b})})\hat{P}_{\vec{a}}\mid\abs{\vec{a}}=\nu+1\}.\label{set:omegaplus}
\end{align}
Let $e_\pm^{(\textrm{ph}, \nu)}$ be the upper bound on the phase error rate when we employ $\Omega_\pm^{(\nu)}(\lambda)$, respectively. $e_\pm^{(\textrm{ph}, \nu)}$ is given by
\begin{align}
e_\pm^{(\textrm{ph}, \nu)}=\min_{\lambda\geq0}\{\lambda e^{(\textrm{b}, \nu)}+\Omega_\pm^{(\nu)}(\lambda)\}.\label{ineq:ephplusminus-bound}
\end{align}
Regarding $\Omega_+^{(\nu)}(\lambda)$, we have an analytical formula that is given in the following lemma (see Appendix~\ref{append:lemma2} for its proof).
\begin{lem}
If $\nu\leq\abs{\mathcal{R}}/2$,
\begin{align}
\Omega_+^{(\nu)}(\lambda)=\frac{1-\lambda}{2}+\nu\frac{1+\lambda}{2\abs{\mathcal{R}}}\label{eq:omegaplus-estimated}
\end{align}
holds for arbitrary $\lambda\geq0$.
\label{lem:omegaplus-estimated}
\end{lem}
By applying this lemma to Eq.~\eqref{ineq:ephplusminus-bound}, we obtain the analytical solution for the upper bound on $e^{(\textrm{ph},\nu)}_+$.
\begin{align}
e_+^{(\textrm{ph}, \nu)}&\leq\min_{\lambda\geq0}\left\{\lambda e^{(\textrm{b}, \nu)}+\frac{1-\lambda}{2}+\nu\frac{1+\lambda}{2\abs{\mathcal{R}}}\right\}\nonumber\\
&=\frac{\abs{\mathcal{R}}+\nu}{2\abs{\mathcal{R}}}+\min_{\lambda\geq0}\left\{\lambda \left(e^{(\textrm{b}, \nu)}-\frac{\abs{\mathcal{R}}-\nu}{2\abs{\mathcal{R}}}\right)\right\}\nonumber\\
&\left\{
\begin{array}{l}
\leq0\quad\displaystyle\hfill\textrm{if }e^{(\textrm{b}, \nu)}<\frac{\abs{\mathcal{R}}-\nu}{2\abs{\mathcal{R}}}=\frac{1}{2}-\frac{\nu}{2\abs{\mathcal{R}}},\\
=\displaystyle\frac{\abs{\mathcal{R}}+\nu}{2\abs{\mathcal{R}}}=\frac{1}{2}+\frac{\nu}{2\abs{\mathcal{R}}}\\
\displaystyle\hfill\textrm{if }e^{(\textrm{b}, \nu)}\geq\frac{\abs{\mathcal{R}}-\nu}{2\abs{\mathcal{R}}}=\frac{1}{2}-\frac{\nu}{2\abs{\mathcal{R}}},
\end{array}
\right.\nonumber\\
\label{ineq:ephplusminus-bound-detail}
\end{align}
where Eq.~\eqref{ineq:ephplusminus-bound-detail} is independent of the length $L$ of one block.
Note that if $\nu>\abs{\mathcal{R}}/2$, Lemma~\ref{lem:omegaplus-estimated} cannot be applied, however, $\Omega_+^{(\nu)}(\lambda)$ for this case is also easily derived by following the same discussion of the derivation of Eq.~\eqref{eq:omegaplus-estimated} (the discussion in Appendix~\ref{append:lemma2} covers this situation, for example $\abs{\mathcal{R}}=\nu=2$).
On the other hand, to derive the upper bound $\Omega_-^{(\nu)}(\lambda)$ in Eq.~\eqref{set:omegaminus}, we need to solve the largest eigenvalue of $\hat{\Pi}_{\vec{a}}^{(\textrm{ph})}-\lambda\hat{\Pi}^{(\textrm{b})}$, which is written as
\begin{align}
&\hat{\Pi}_{\vec{a}}^{(\textrm{ph})}-\lambda\hat{\Pi}^{(\textrm{b})}\nonumber\\
\begin{split}
=&\sum_{m=1}^L\hat{P}\left(\ket{m}_\textrm{B}\right)\left(\frac{1}{2}\delta_{a_m, 1}+\frac{1}{2\abs{\mathcal{R}}}\sum_{n:\abs{m-n}\in\mathcal{R}}\delta_{a_n, 1}\right)\\
&-\lambda\left(\frac{1}{2}\hat{1}_\textrm{B}-\frac{1}{2\abs{\mathcal{R}}}\sum_{(m, n):\abs{m-n}\in\mathcal{R}}\ket{m}_\textrm{B}{}_\textrm{B}\bra{n}\right)
\end{split}\nonumber\\
=&\sum_{m=1}^L\hat{P}\left(\ket{m}_\textrm{B}\right)\left(\frac{\delta_{a_m, 1}-\lambda}{2}+\frac{1}{2\abs{\mathcal{R}}}\sum_{n:\abs{m-n}\in\mathcal{R}}\delta_{a_n, 1}\right)\nonumber\\
&+\frac{\lambda}{2\abs{\mathcal{R}}}\sum_{(m, n):\abs{m-n}\in\mathcal{R}}\ket{m}_\textrm{B}{}_\textrm{B}\bra{n}.
\label{eq:operator-omegaminus}
\end{align}
To discuss the eigenvalues of Eq.~\eqref{eq:operator-omegaminus}, we recall the fact that the translation operation $\hat{V}_{\vec{a}, \vec{a}'}$ defined for $(\vec{a}, \vec{a}')$ such that $a_k=a'_{k+_L\kappa}$ is satisfied for any $k\in\{1,\dots, L\}$ with specific $\kappa\in\{1,\dots,L\}$,
\begin{align}
\hat{V}_{\vec{a}, \vec{a}'}:=\sum_{m=1}^L\ket{m+_L\kappa}_\textrm{B}{}_\textrm{B}\bra{m}\label{defi:translation-operator}
\end{align}
does not change the eigenvalues.
Hence, the eigenvalues of Eq.~\eqref{eq:operator-omegaminus} with $\vec{a}$ and $\vec{a}'$ are the same if there exists $\kappa$ ($1\leq \kappa\leq L$) such that $a_k=a'_{k+_L\kappa}$ is satisfied for any $k\in\{1,\dots, L\}$.
By using this, for $\nu=2$, it is enough to consider the case $\vec{a}=(\overbrace{0\dots0}^{\abs{\mathcal{R}}/2}10\dots0)$, where the matrix representation of Eq.~\eqref{eq:operator-omegaminus} is written as
\begin{align}
\frac{1}{2\abs{\mathcal{R}}}\mathrm{diag}\{\overbrace{1,\dots,1}^{\abs{\mathcal{R}}/2}, \abs{\mathcal{R}}, \overbrace{1,\dots,1}^{\abs{\mathcal{R}}/2}, 0\dots0\}_L\nonumber\\
\hfill-\frac{\lambda}{2}I_L+\frac{\lambda}{2\abs{\mathcal{R}}}A^{(\abs{\mathcal{R}}/2)}_L.\label{matrix:omegaminus_nu=2}
\end{align}
Here, $A^{(k)}_n$ is an $n\times n$ matrix satisfying
\begin{align}
(A^{(k)}_n)_{l, m}=
\begin{cases}
1&\textrm{if }1\leq\abs{l-m}\leq k,\\
1&\textrm{if }n-k\leq\abs{l-m}\leq n-1,\\
0&\textrm{otherwise}.
\end{cases}
\end{align}
In order to obtain the largest eigenvalue of Eq.~\eqref{matrix:omegaminus_nu=2} for large $L$, we take a numerical approach. For $\nu=1$, however, the upper bound on $\Omega_-^{(\nu)}(\lambda)$ can be easily derived, and we obtain the following theorem for the upper bound on $e^{(\textrm{ph}, 1)}$ (see Appendix~\ref{append:theorem1} for its proof).
\begin{theo}
For $0\leq e^{(\mathrm{b}, 1)}\leq\frac{\abs{\mathcal{R}}-1}{2\abs{\mathcal{R}}}$,
\begin{align}
e^{(\mathrm{ph}, 1)}\leq\frac{\abs{\mathcal{R}}+1}{\abs{\mathcal{R}}-1}e^{(\mathrm{b}, 1)}
\label{eq:phase_error_nu=1}
\end{align}
holds. Also, for $e^{(\mathrm{b}, 1)}\geq\frac{\abs{\mathcal{R}}-1}{2\abs{\mathcal{R}}}$, we have
\begin{align}
e^{(\mathrm{ph}, 1)}\leq\frac{1}{2}+\frac{1}{2\abs{\mathcal{R}}}.
\end{align}
\label{theo:phase_error_nu=1}
\end{theo}
For $\nu>1$, the upper bound on $e^{(\textrm{ph}, \nu)}$ is derived as the maximum value of the convex combination of the one on $e_+^{(\textrm{ph}, \nu)}$ and $e_-^{(\textrm{ph}, \nu)}$ as
\begin{align}
e^{(\textrm{ph}, \nu)}\leq\max_{\substack{p: 0\leq p\leq1\\px_++(1-p)x_-=e^{(\textrm{b}, \nu)}}}\left\{pf^{(\nu)}_+(x_+)+(1-p)f^{(\nu)}_-(x_-)\right\},\label{ineq:eph-mixture}
\end{align}
where $f^{(\nu)}_\pm(x)$ is defined by
\begin{align}
f^{(\nu)}_\pm(x)=\min_{\lambda_\pm\geq0}\left\{\lambda_\pm x+\Omega_\pm^{(\nu)}(\lambda_\pm)\right\}.
\end{align}

\begin{figure}[t]
\begin{center}
\begin{tabular}{c}

\begin{minipage}{1\hsize}
\begin{center}
\includegraphics[height=4.6cm]{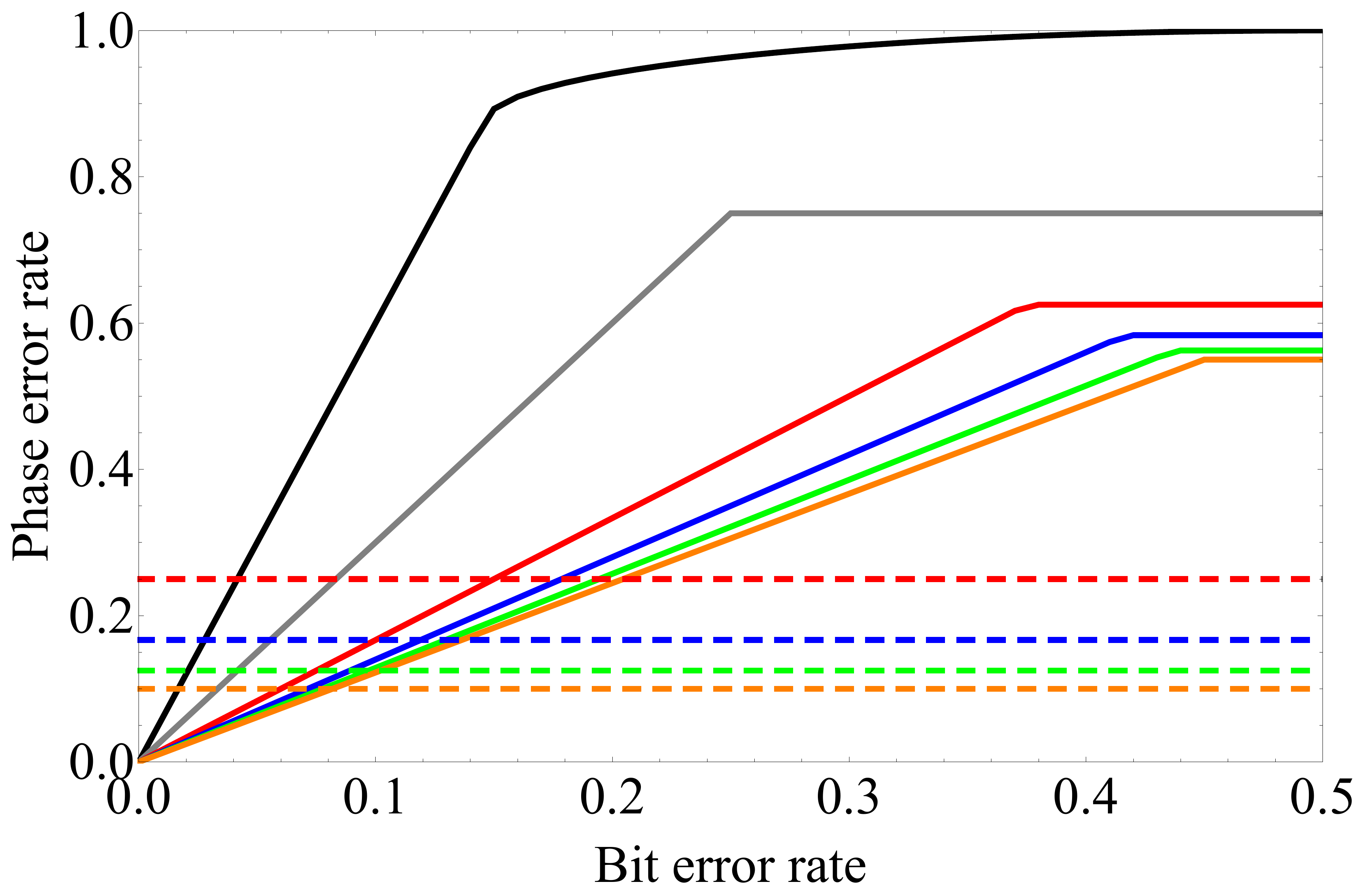}
\\
\hspace{0.3cm} (a) $\nu=1$
\end{center}
\end{minipage}\\

\begin{minipage}{1\hsize}
\begin{center}
\includegraphics[height=4.62cm]{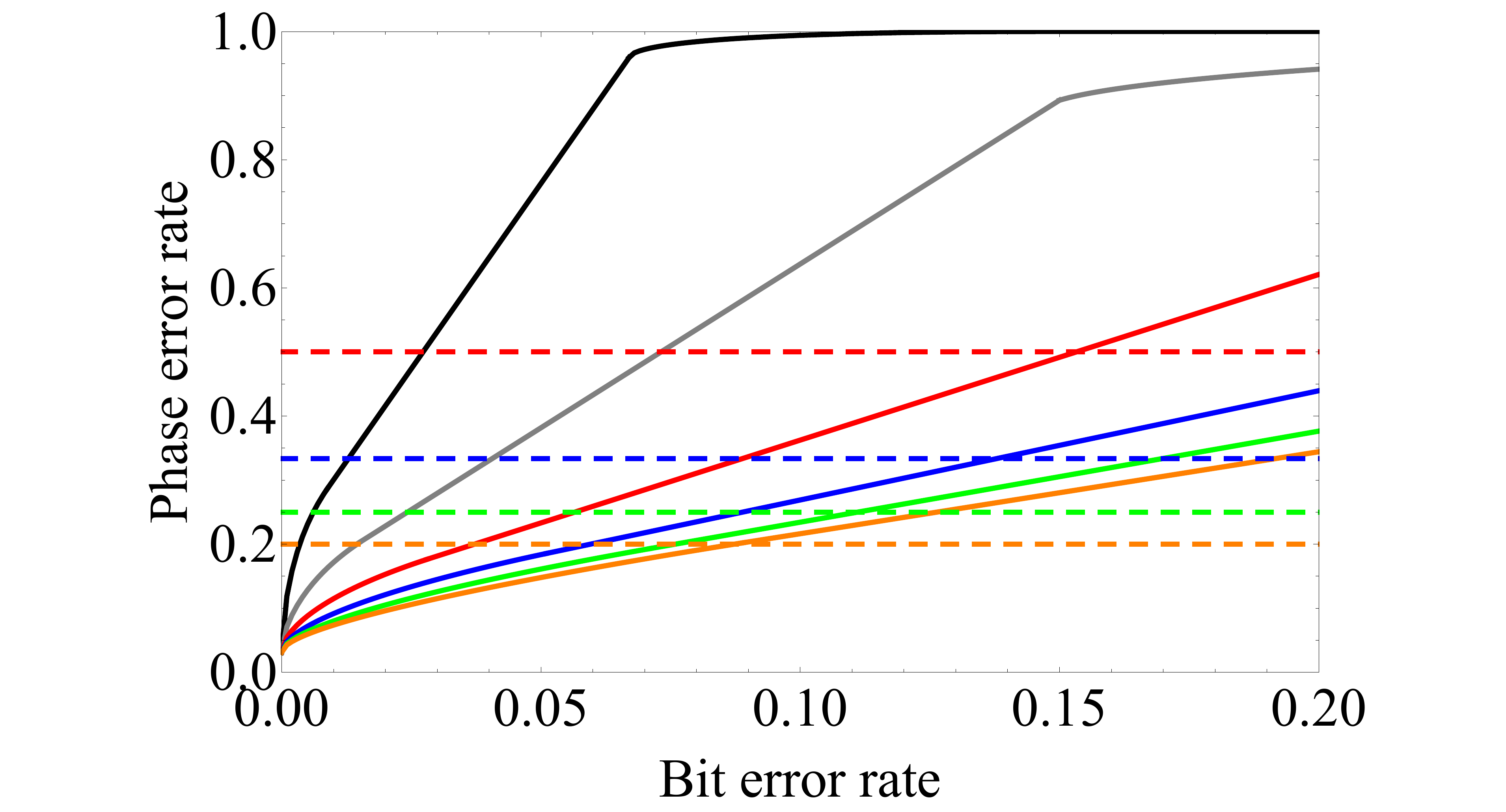}
\\
\hspace{0.3cm} (b) $\nu=2$
\end{center}
\end{minipage}

\end{tabular}
\caption{(Color online) The upper bound on the phase error rate $e^{(\textrm{ph}, \nu)}$ as a function of the bit error rate $e^{(\textrm{b}, \nu)}$ in the case of (a) $\nu=1$ and (b) $\nu=2$, respectively.
From bottom to top, the solid lines correspond to the case of the SNRDPS protocol with $(L, \abs{\mathcal{R}})=(32, 10), (32, 8), (32, 6), (32, 4)$ and $(32, 2)$, respectively.
The black solid line on the topmost of both figures corresponds to the case for the original DPS protocol~\cite{tamaki2012unconditional} with $L=32$.
Also, from bottom to top, the dashed lines correspond to the case of the RRDPS protocol~\cite{sasaki2014practical} with $(L, \abs{\mathcal{R}})=(11, 10), (9, 8), (7, 6)$ and $(5, 4)$, respectively.}
\label{fig:eph-1and2}
\end{center}
\end{figure}

In FIG.~\ref{fig:eph-1and2} (a) and (b), we plot the estimated upper bound on the phase error rate $e^{(\textrm{ph}, \nu)}$ as a function of $e^{(\textrm{b}, \nu)}$ with the number of delays $\abs{\mathcal{R}}\in\{2,4,6,8,10\}$ for $\nu\in\{1, 2\}$. 
From these figures, even if the number of the random delays is small (say $\abs{\mathcal{R}}=2$), we can observe a significant improvement over the original DPS protocol.
Moreover, when we compare the resulting phase error rate of the SNRDPS  protocol (solid lines) with the one of the RRDPS protocol (dashed lines) with the small random delays, the phase error rate of the SNRDPS protocol is smaller if the bit error rate is small. Here, we assume the phase error rate of the RRDPS protocol as $e^{(\textrm{ph})}=\nu/\abs{\mathcal{R}}$~\cite{sasaki2014practical}.

\section{Key generation rates}
\label{sec:key-generation-rates}
\begin{figure}[t]
\begin{center}
\begin{tabular}{c}

\begin{minipage}{1\hsize}
\begin{center}
\includegraphics[height=5cm]{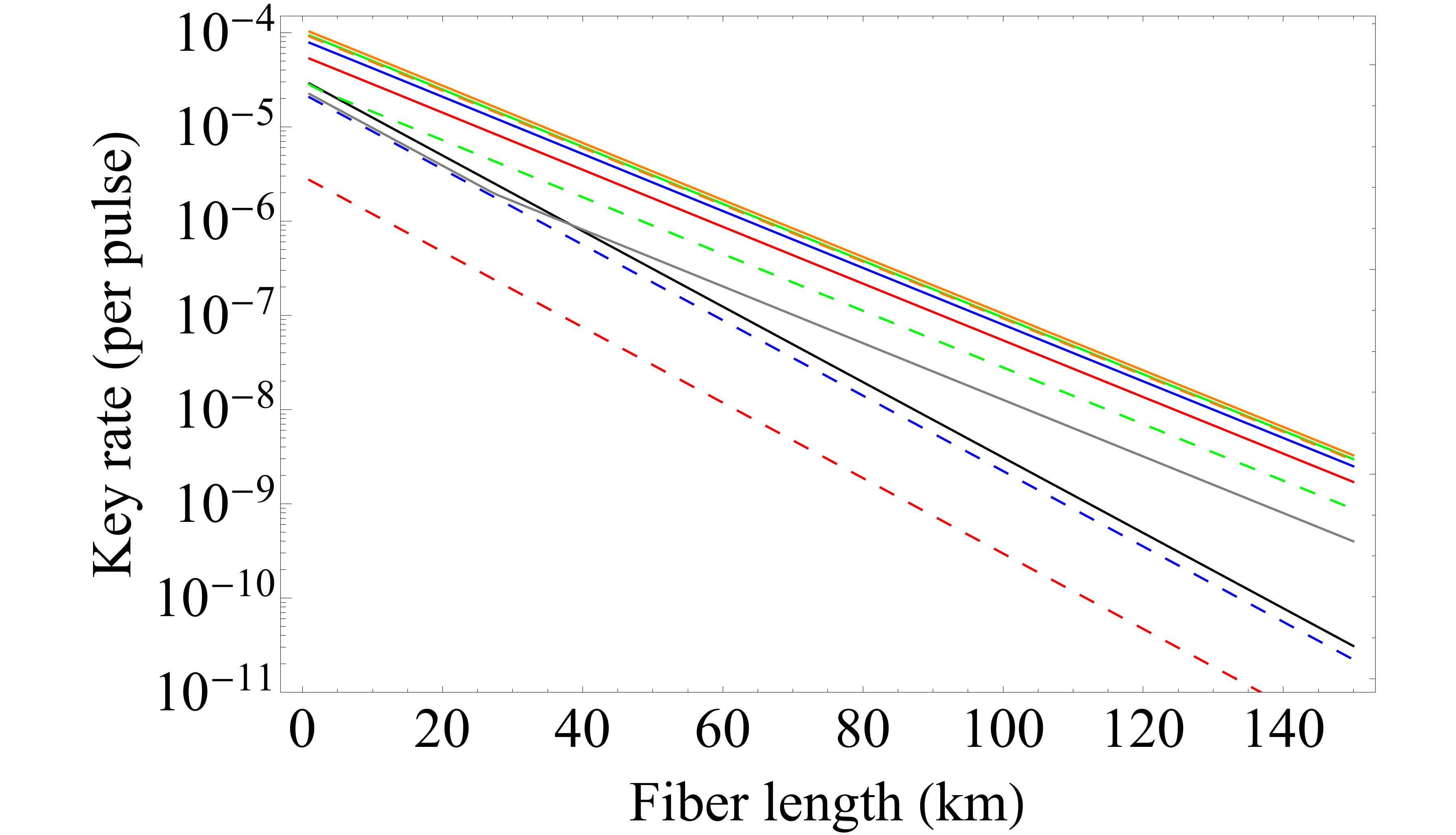}\\
\hspace{0.5cm} (a) $e^{(\textrm{b})}=2\%$
\vspace{0.3cm}
\end{center}
\end{minipage}\\

\begin{minipage}{1\hsize}
\begin{center}
\includegraphics[height=5cm]{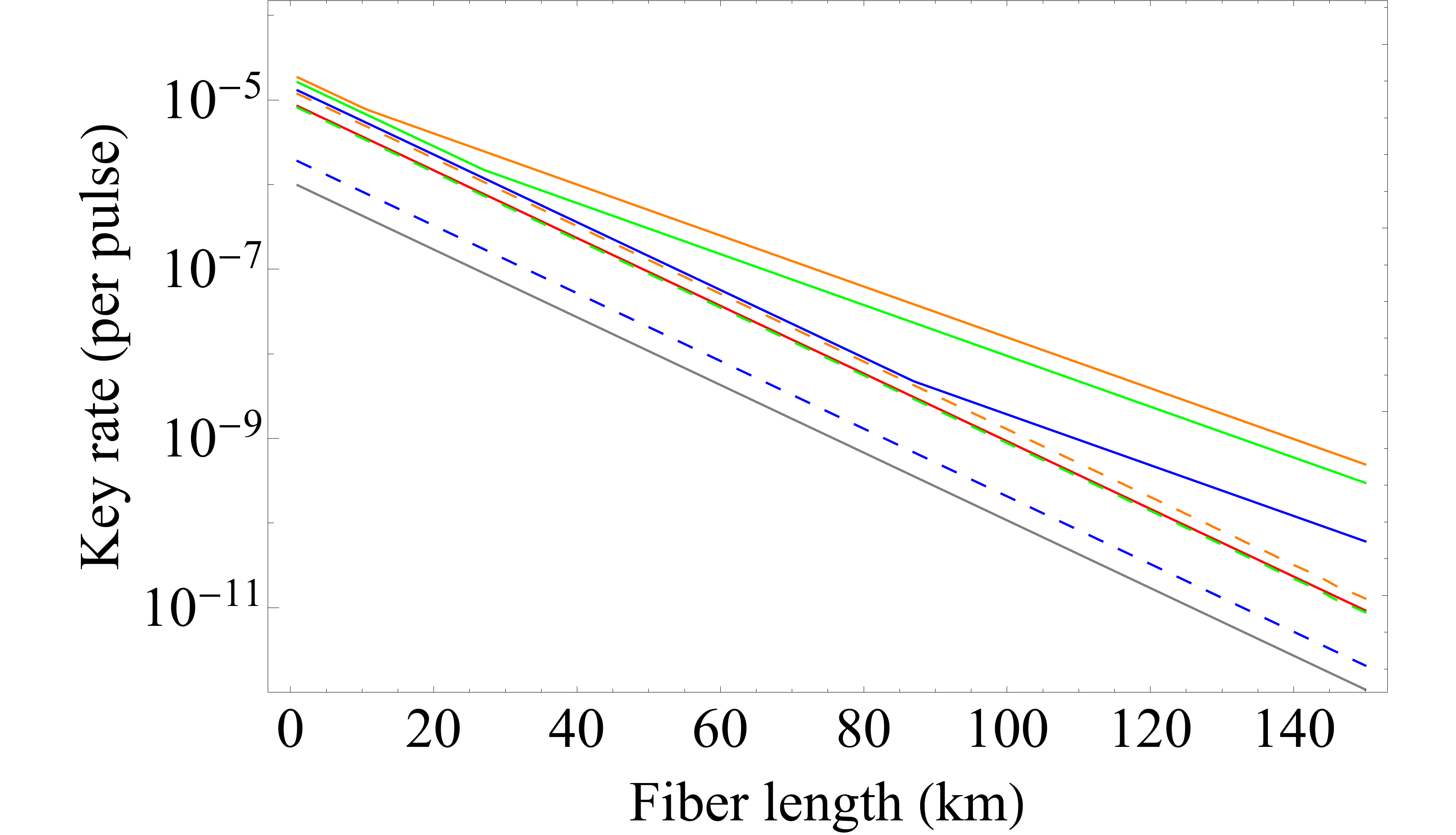}\\
\hspace{0.5cm} (b) $e^{(\textrm{b})}=5\%$
\end{center}
\end{minipage}

\end{tabular}
\caption{(Color online) The key generation rate $G$ in logarithmic scale vs fiber length. 
We assume Bob's detection efficiency of $10\%$, the channel transmittance of 0.2dB/km, and the bit error rate $e^{(\textrm{b})}=2\%$ in (a) and $e^{(\textrm{b})}=5\%$ in (b), respectively.
In this simulation, we have optimized the mean photon number $\mu=\abs{\alpha}^2$.
From top to bottom, the solid lines are for the SNRDPS protocols with $(L, \abs{\mathcal{R}})=(32, 10), (32, 8), (32, 6), (32, 4)$ and $(32, 2)$, respectively.
The key generation rate of the original DPS protocol~\cite{tamaki2012unconditional} with $L=32$ is plotted in the bottommost solid line in (a).
In the original DPS protocol, the secret key cannot be extracted when $e^{(\textrm{b})}=5\%$. 
Also, from top to bottom, the dashed lines express the resulting key generation rates for the RRDPS protocol with $(L, \abs{\mathcal{R}})=(11, 10), (9, 8), (7, 6)$ and $(5, 4)$, respectively.
Note that the dashed lines corresponding $(L, \abs{\mathcal{R}})=(5, 4)$ disappeared from (b) since the rate is zero.
}
\label{fig:keyrate}
\end{center}
\end{figure}

In this section, we present our main results, namely, the key generation rate of the SNRDPS protocol is significantly enhanced over the one of the original DPS protocol only by employing a few additional delays such as $\abs{\mathcal{R}}=2$.
In FIG.~\ref{fig:keyrate}, we compare the key generation rate $G$ per pulse in Eq.~\eqref{eq:keyrate} for three protocols:
(i) the original DPS protocol~\cite{tamaki2012unconditional}, 
(ii) the RRDPS protocol~\cite{sasaki2014practical} without monitoring the disturbance when Bob employs the number of random delays $\abs{\mathcal{R}}$, 
and (iii) the SNRDPS protocol when Bob employs the number of random delays $\abs{\mathcal{R}}$.
From FIG.~\ref{fig:keyrate} (a), we can see that the key generation rate of the SNRDPS protocol with $\abs{\mathcal{R}}=2$ outperforms the original DPS protocol when the fiber length is more than about 40km and the bit error rate $e^{(\textrm{b})}$ is $2\%$.
Also, the SNRDPS protocol always outperforms the original DPS protocol when $\abs{\mathcal{R}}\geq4$ and $e^{(\textrm{b})}=2\%$.
Moreover, from FIG.~\ref{fig:keyrate} (b), the SNRDPS protocol provides a positive key generation rate even though the original DPS protocol cannot generate the secret key.
Also, in both figures, by comparing two lines with the same colors, we confirm that the SNRDPS protocol outperforms the RRDPS protocol with the same $\abs{\mathcal{R}}$ up to $\abs{\mathcal{R}}=10$.
This means that, if the amount of randomness is small as $2\leq\abs{\mathcal{R}}\leq 10$ and $e^{(\textrm{b})}\leq 5\%$, the key generation rate of the SNRDPS protocol outperforms the one of the RRDPS protocol when both protocols employ the same number of random delays $\abs{\mathcal{R}}$ and Alice and Bob do not monitor the bit error rate in the RRDPS protocol.

\begin{figure}[t]
\begin{center}
\begin{tabular}{c}

\begin{minipage}{1\hsize}
\begin{center}
\includegraphics[height=5cm]{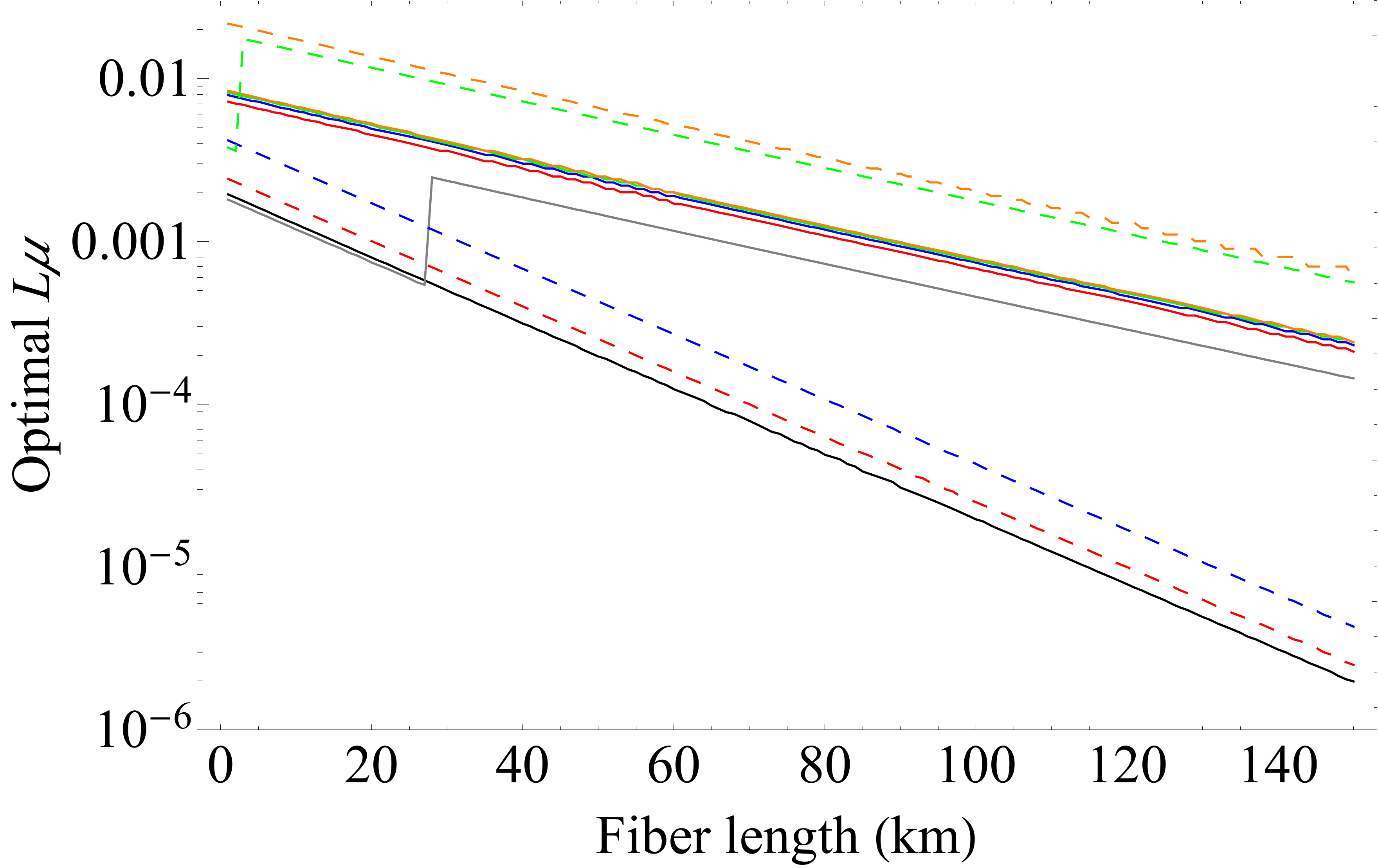}\\
\hspace{0.5cm} (a) $e^{(\textrm{b})}=2\%$\\
\vspace{0.3cm}
\end{center}
\end{minipage}\\

\begin{minipage}{1\hsize}
\begin{center}
\includegraphics[height=5cm]{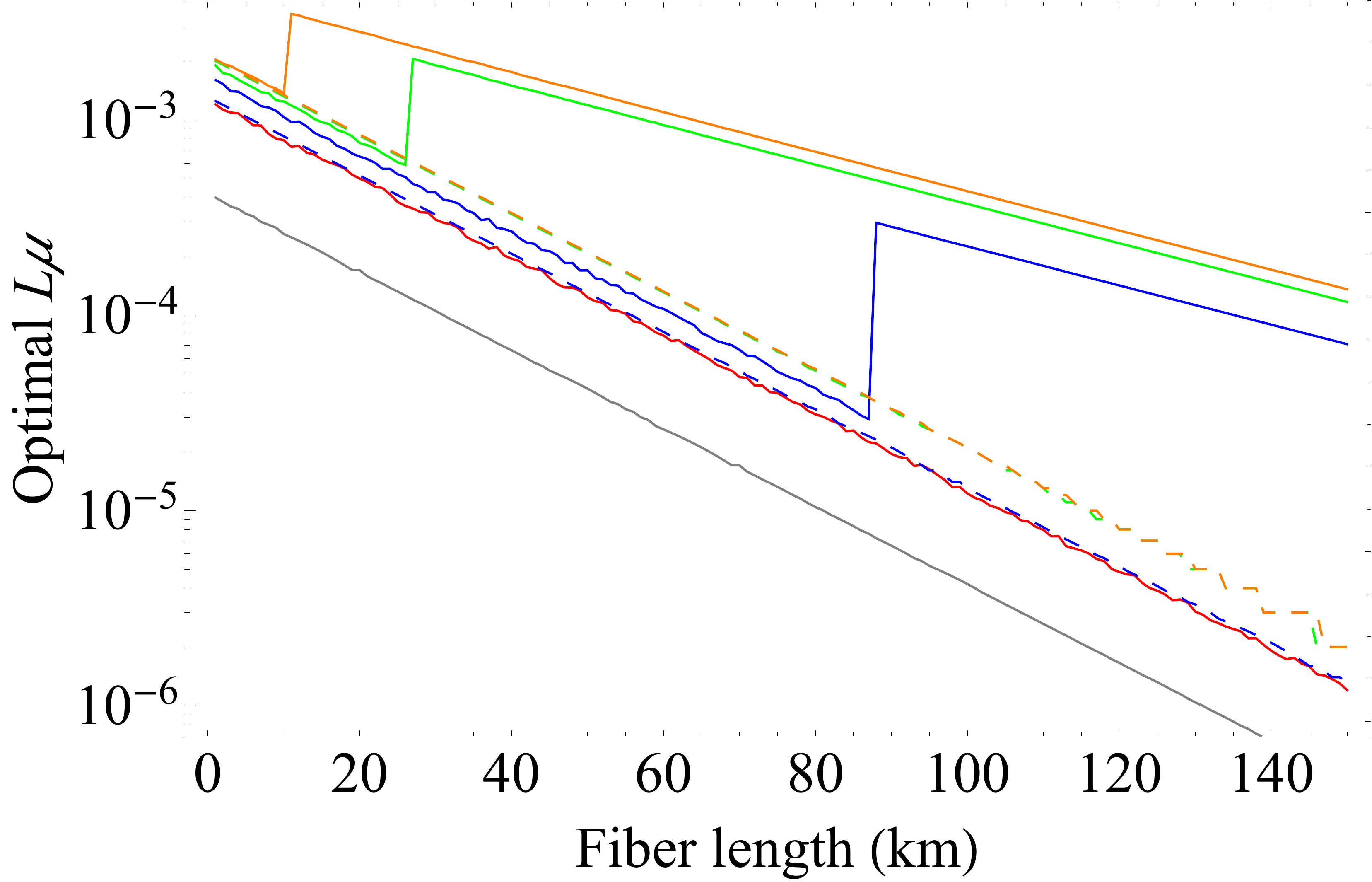}\\
\hspace{0.5cm} (b) $e^{(\textrm{b})}=5\%$
\end{center}
\end{minipage}

\end{tabular}
\caption{(Color online) The optimal light intensity $L\mu$ to achieve the key generation rate in FIG.~\ref{fig:keyrate} in logarithmic scale vs fibre length when the bit error rate $e^{(\textrm{b})}=2\%$ in (a) and $e^{(\textrm{b})}=5\%$ in (b).
From top to bottom, the solid lines are for the cases of the SNRDPS protocol with $(L, \abs{\mathcal{R}})=(32, 10), (32, 8), (32, 6), (32, 4)$, and $(32, 2)$, respectively.
The light intensity of the original DPS protocol~\cite{tamaki2012unconditional} with $L=32$ is plotted in the bottom solid line in (a).
In the original DPS protocol, the optimal light intensity cannot be found when $e^{(\textrm{b})}=5\%$.
Also, from top to bottom, the dashed lines express the cases for the RRDPS protocol with $(L, \abs{\mathcal{R}})=(11, 10), (9, 8), (7, 6)$, and $(5, 4)$, respectively.
Note that the dashed lines corresponding $(L, \abs{\mathcal{R}})=(5, 4)$ is disappeared in (b) since the optimal light intensity cannot be found.
}
\label{fig:light_intensity}
\end{center}
\end{figure}

Next, we discuss the transmittance dependency of the key generation rates for the SNRDPS protocol. 
For this, we assume a fibre-based QKD system, and the detection efficiency $Q$ is assumed to be $Q=\frac{L\mu\eta}{2}\mathrm{e}^{-L\mu\eta}$.
Here, $\mu:=\abs{\alpha}^2$ denotes the mean photon number of per sending pulse, and
$\eta=\eta(l)$ denotes channel transmittance with the fibre length $l$ as 
$\eta(l):=\eta_0\times10^{-\frac{0.2 l}{10}}$ with $\eta_0$ denoting Bob's detection efficiency. 
If $\mu$ is small ($\mu<1$), $Q$ and $p_\nu$ are approximated to $Q\sim\mathcal{O}(\mu\eta)$ and $p_\nu\sim\mathcal{O}(\mu^\nu)$, respectively.
Suppose that Alice and Bob generate the secret key from the $\nu$-photon emissions. 
In this case, by considering Eve's attack, the total detection efficiency $Q\sim\mathcal{O}(\mu\eta)$ minus the probability of emitting more than $\nu+1$ photons (approximated to $\mu^{\nu+1}$) has to be positive, resulting in $\mathcal{O}(\mu^{\nu+1})\leq\mathcal{O}(\mu\eta)$.
From this, we obtain the dependency of $\mu$ over the transmittance as
\begin{align}
\mu\sim\mathcal{O}(\eta^{1/\nu}),\label{eq:order_mu}
\end{align}
and hence the key generation rate $G$ behaves as
\begin{align}
G\sim Q\sim\mathcal{O}(\eta^{(\nu+1)/\nu}).\label{eq:order_G}
\end{align}

In FIG.~\ref{fig:keyrate}~(a), all the lines of the SNRDPS protocol except for the one with $(L, \abs{\mathcal{R}})=(32, 2)$ and the one of the RRDPS protocol with $\abs{\mathcal{R}}=10$ result in the transmittance dependency as $G\sim\mathcal{O}(\eta^{3/2})$, which means that the secret key is extracted from the two-photon emission events in addition to the single-photon emission events.
On the other hand, the line of the original DPS protocol and the ones of the RRDPS protocol with $\abs{\mathcal{R}}=4, 6$ result in the transmittance dependency as $G\sim\mathcal{O}(\eta^2)$ since the upper bound on the phase error rate of the two-photon emission events is too high to extract the secret key.
Moreover, the SNRDPS protocol with $(L, \abs{\mathcal{R}})=(32, 2)$ and the RRDPS protocol with $\abs{\mathcal{R}}=8$ provide the key generation rate of the form $G\sim\mathcal{O}(\eta^{2})$ for shorter distance and $G\sim\mathcal{O}(\eta^{3/2})$ for longer distance.
The implication of this is that when the loss increases, the two-photon contribution becomes larger, and moreover the bit error rate of $2\%$ is still small enough to generate the key from the two-photon emission event.
Also, in FIG.~\ref{fig:keyrate}~(b), all the lines of the SNRDPS protocol except the one with $(L, \abs{\mathcal{R}})=(32, 4)$ and $(32, 2)$ result in the transmittance dependency as $G\sim\mathcal{O}(\eta^{2})$ for shorter distance and $G\sim\mathcal{O}(\eta^{3/2})$ for longer distance, while all the remaining lines result in the transmittance dependency as $G\sim\mathcal{O}(\eta^2)$.
Even when the bit error rate is $5\%$, the properties of transmittance dependency of the key generation rates including the change of the scaling from the short and the long distance regimes can be explained with the same reason as the ones in Fig 5 (a).

Finally, FIG.~\ref{fig:light_intensity}~(a) and (b) show the optimal mean photon number $L\mu$ to realize the key generation rates in FIG.~\ref{fig:keyrate}~(a) and (b), respectively.
For all the protocols, it can be found that the optimal mean photon number scales as $\mathcal{O}(\eta^{1/\nu})$ for $\nu$-photon emission event, which is the same scaling as Eq.~\eqref{eq:order_mu}.
Also, the discontinuous point of the lines in FIG.~\ref{fig:light_intensity}~(a) and (b), which represents the boundary of the presence or absence of the two-photon contribution, corresponds to the changing point of the scaling of the key generation rates in FIG.~\ref{fig:keyrate}~(a) and (b), respectively.

\section{conclusion}
\label{sec:conclusion}
In conclusion, in this paper, we have proposed a new DPS-type QKD protocol with a small random delays at Bob's measurement and analyzed its information-theoretical security.
For this protocol, we have estimated an upper bound on the phase error rate for Alice's single and two-photon emission events by using the bit error rate information.
Besides, we have simulated and compared the key generation rates for the SNRDPS protocol with $\abs{\mathcal{R}}\in\{2,4,6,8,10\}$, the one for the original DPS protocol, and the ones for the RRDPS protocol with $\abs{\mathcal{R}}\in\{4,6,8,10\}$.
As a result, we found that the performance of the SNRDPS protocol is significantly enhanced from the original DPS protocol even when Bob employs only a few number of delays such as $\abs{\mathcal{R}}=2$.
Moreover, we found that if $\abs{\mathcal{R}}\leq 10$, the key generation rate of the SNRDPS protocol based on our analysis outperforms the RRDPS protocol without monitoring the disturbance~\cite{sasaki2014practical} when the same number of random delays is employed.

The SNRDPS protocol is an intermediate protocol between the original DPS and the RRDPS protocols in terms of the practicality and bit error tolerance, and this protocol increases the variety of future implementation for the DPS-type QKD protocol.

\acknowledgments
We thank Toshihiko Sasaki, Hoi-Kwong Lo, Koji Azuma, Rikizo Ikuta, and Yuki Takeuchi for helpful discussions. 
NI acknowledges support from the MEXT/JSPS KAKENHI Grant Number 16H02214.
This work is in part funded by ImPACT Program of Council for Science, Technology and Innovation (Cabinet Office, Government of Japan).


\appendix
\section{Proof of Lemma~\ref{fact:dial-actual-relation}}
\label{append:dial-measurement}
Here, we prove Lemma~\ref{fact:dial-actual-relation} in the main text.
First, we have that $\hat{E}_{k, s}^{(r)}$ satisfies the following relation
\begin{align}
\hat{E}_{k, s}^{(r)}=
\begin{cases}
\hat{B}_{k, s}^{(r)}&~~~\textrm{if }1\leq k\leq L-r,\\
\hat{B}_{k-L+r, s}^{(L-r)}&~~~\textrm{if }L-r+1\leq k\leq L.
\end{cases}
\label{eq:E-B_POVM_relation}
\end{align}
This is so because $\hat{E}_{k, s}^{(r)}$ is written as
\begin{align}
\hat{E}_{k, s}^{(r)}=\frac{1}{2}\hat{P}\left(\frac{\ket{k}_\textrm{B}+(-1)^s\ket{k+r}_\textrm{B}}{\sqrt{2}}\right)=\hat{B}_{k, s}^{(r)}
\end{align}
for $1\leq k\leq L-r$ and 
\begin{align}
\hat{E}_{k, s}^{(r)}=&\frac{1}{2}\hat{P}\left(\frac{\ket{k}_\textrm{B}+(-1)^s\ket{k-L+r}_\textrm{B}}{\sqrt{2}}\right)\nonumber\\
=&\frac{1}{2}\hat{P}\left(\frac{(-1)^s\ket{k-L+r+(L-r)}_\textrm{B}+\ket{k-L+r}_\textrm{B}}{\sqrt{2}}\right)\nonumber\\
=&\hat{B}_{k-L+r, s}^{(L-r)}
\end{align}
for $L-r+1\leq k\leq L$.
By using Eq.~\eqref{eq:E-B_POVM_relation} and regarding $\hat{B}_{k, s}^{(r)}=0$ if $k\leq0$ or $L-r+1\leq k$, we have the following equation.
\begin{align}
\hat{E}_{k, s}^{(r)}=\hat{B}_{k, s}^{(r)}+\hat{B}_{k-L+r, s}^{(L-r)}.\label{eq:e-b_measurement}
\end{align}
If we fix the delay of the dial measurement as $r'$, the probability that Bob obtains the bit value $s$ and announces the pair of integers $(i, j)~~(i<j)$ is given by
\begin{align}
\cpr{s\wedge(i, j)}{r'}_\textrm{dial}=&\tr{\hat{\rho}\hat{E}_{i, s}^{(r')}}\delta_{j-i, r'}\nonumber\\
&+\tr{\hat{\rho}\hat{E}_{j, s}^{(r')}}\delta_{j-i, L-r'}.\label{eq:cpr-dial}
\end{align}
To simulate the dial measurement with the delay $r'$ by using Bob's actual measurement, he randomly switches the delays of the actual measurement $r'$ and $L-r'$. The probability that Bob obtains the outcome $s$ and announces $(i, j)$ when he performs the actual measurement is written as
\begin{align}
\tr{\hat{\rho}\hat{B}_{i, s}^{(r')}}\delta_{j-i, r'}
\end{align}
if the delay is $r'$ and
\begin{align}
\tr{\hat{\rho}\hat{B}_{i, s}^{(L-r')}}\delta_{j-i, L-r'}
\end{align}
if the delay is $L-r'$. Let us define $\cpr{s\wedge(i,j)}{r\in\{r', L-r'\}}_\textrm{actual}$ as the probability that Bob obtains $s$ and announces $(i, j)$ when he performs the actual measurement with the delay $r=r'$ or $r=L-r'$ uniformly at random. $\cpr{s\wedge(i, j)}{r\in\{r', L-r'\}}_\textrm{actual}$ is written as
\begin{align}
&\cpr{s\wedge(i, j)}{r\in\{r', L-r'\}}_\textrm{actual}\nonumber\\
=&\frac{1}{2}\tr{\hat{\rho}\hat{B}_{i, s}^{(r')}}\delta_{j-i, r'}+\frac{1}{2}\tr{\hat{\rho}\hat{B}_{i, s}^{(L-r')}}\delta_{j-i, L-r'}\nonumber\\
=&\frac{1}{2}\tr{\hat{\rho}\hat{B}_{i, s}^{(r')}}\delta_{j-i, r'}+\frac{1}{2}\tr{\hat{\rho}\hat{B}_{j-L+r', s}^{(L-r')}}\delta_{j-i, L-r'}\nonumber\\
\begin{split}
=&\frac{1}{2}\tr{\hat{\rho}\left[\hat{E}_{i, s}^{(r')}-\hat{B}_{i-L+r', s}^{(L-r')}\right]}\delta_{j-i, r'}\\
&+\frac{1}{2}\tr{\hat{\rho}\left[\hat{E}_{j, s}^{(r')}-\hat{B}_{j, s}^{(r')}\right]}\delta_{j-i, L-r'}
\end{split}\nonumber\\
\begin{split}
=&\frac{1}{2}\left[\tr{\hat{\rho}\hat{E}_{i, s}^{(r')}}\delta_{j-i, r'}+\tr{\hat{\rho}\hat{E}_{j, s}^{(r')}}\delta_{j-i, L-r'}\right]\\
&-\frac{1}{2}\tr{\hat{\rho}\left[\hat{B}_{i-L+r', s}^{(L-r')}\delta_{j-i, r'}+\hat{B}_{j, s}^{(r')}\delta_{j-i, L-r'}\right]}
\end{split}\nonumber\\
=&\frac{1}{2}\cpr{s\wedge(i, j)}{r'}_\textrm{dial},
\end{align}
where we have used Eqs.~\eqref{eq:e-b_measurement} and~\eqref{eq:cpr-dial} in the third and fifth equalities, respectively. Also, we have used
\begin{align}
&\hat{B}_{i-L+r', s}^{(L-r')}\delta_{j-i, r'}+\hat{B}_{j, s}^{(r')}\delta_{j-i, L-r'}\nonumber\\
=&\hat{B}_{i-L+r', s}^{(L-r')}\delta_{i-L+r', j-L}+\hat{B}_{j, s}^{(r')}\delta_{j, L-r'+i}\nonumber\\
=&\hat{B}_{j-L, s}^{(L-r')}\delta_{i-L+r', j-L}+\hat{B}_{L-r'+i, s}^{(r')}\delta_{j, L-r'+i}=0
\label{eq:actual_measurement_vanishing}
\end{align}
in the fifth equality, which is satisfied since $\hat{B}_{k, s}^{(r)}=0$ for $k\leq0 $ or $L-r+1\leq k$, and $(i, j)$ satisfies $1\leq i<j\leq L$.

\section{Detail of calculation of bit and phase error POVMs} 
\label{append:bit-phase-error-POVMs}
Here, we detail the calculation of the equations of bit and phase error POVMs. First, Eq.~\eqref{defi:ephPOVM_ij} is derived as follows.
\begin{widetext}
\begin{align}
\hat{e}_{i, j}^{(\textrm{ph})}:=&\sum_{s=0}^1\sum_{k=1}^3\hat{M}_k^{(i, j)\dagger}\hat{P}\left(\hat{H}\ket{s}_\textrm{Aq}\right)\hat{M}_k^{(i, j)}\otimes\frac{1}{\abs{\mathcal{R}}}\sum_{r\in\mathcal{R}}\left[\hat{F}_i^{(r)\dagger}\hat{P}\left(\hat{H}\ket{\bar{s}}_\textrm{B}\right)\hat{F}_i^{(r)}\delta_{j, i+r}+\hat{F}_j^{(r)\dagger}\hat{P}\left(\hat{H}\ket{\bar{s}}_\textrm{B}\right)\hat{F}_j^{(r)}\delta_{i, j+r-L}\right]\nonumber\\
=&\frac{1}{\abs{\mathcal{R}}}\sum_{s=0}^1\sum_{k=1}^3\hat{M}_k^{(i, j)\dagger}\hat{P}\left(\hat{H}\ket{s}_\textrm{Aq}\right)\hat{M}_k^{(i, j)}\otimes\left[\hat{F}_i^{(j-i)\dagger}\hat{P}\left(\hat{H}\ket{\bar{s}}_\textrm{B}\right)\hat{F}_i^{(j-i)}+\hat{F}_j^{(L-(j-i))\dagger}\hat{P}\left(\hat{H}\ket{\bar{s}}_\textrm{B}\right)\hat{F}_j^{(L-(j-i))}\right]\nonumber\\
=&\frac{2}{\abs{\mathcal{R}}}\sum_{s=0}^1\sum_{k=1}^3\hat{M}_k^{(i, j)\dagger}\hat{P}\left(\hat{H}\ket{s}_\textrm{Aq}\right)\hat{M}_k^{(i, j)}\otimes\hat{\mathcal{F}}_{i, j}^{\dagger}\hat{P}\left(\hat{H}\ket{\bar{s}}_\textrm{B}\right)\hat{\mathcal{F}}_{i, j}\nonumber\\
=&\frac{1}{\abs{\mathcal{R}}}\sum_{s=0}^1\left[\hat{P}\left(\ket{s}_{\textrm{A}, i}\ket{\bar{s}}_{\textrm{A}, j}\right)+\frac{1}{2}\sum_{t=0}^1\hat{P}\left(\ket{t}_{\textrm{A}, i}\ket{t}_{\textrm{A}, j}\right)\right]\otimes\hat{P}\left(\delta_{\bar{s},1}\ket{i}_\textrm{B}+\delta_{\bar{s}, 0}\ket{j}_\textrm{B}\right)\nonumber\\
=&\frac{1}{\abs{\mathcal{R}}}\left[\hat{P}\left(\ket{0}_{\textrm{A}, i}\ket{1}_{\textrm{A}, j}\right)\otimes\hat{P}\left(\ket{i}_\textrm{B}\right)+\hat{P}\left(\ket{1}_{\textrm{A}, i}\ket{0}_{\textrm{A}, j}\right)\otimes\hat{P}\left(\ket{j}_\textrm{B}\right)\right]+\frac{1}{2\abs{\mathcal{R}}}\sum_{t=0}^1\hat{P}\left(\ket{t}_{\textrm{A}, i}\ket{t}_{\textrm{A}, j}\right)\otimes\left[\hat{P}\left(\ket{i}_\textrm{B}\right)+\hat{P}\left(\ket{j}_\textrm{B}\right)\right].
\end{align}
\end{widetext}
Here, we define $\bar{s}:=s\oplus1$ and $\hat{\mathcal{F}}_{i, j}:=\frac{1}{\sqrt{2}}\hat{H}\ket{1}_\textrm{Bq}{}_\textrm{B}\bra{i}+\frac{1}{\sqrt{2}}\hat{H}\ket{0}_\textrm{Bq}{}_\textrm{B}\bra{j}$.

Next, we detail the derivation of Eq.~\eqref{eq:unitary-ebit} as follows. $\hat{\Pi}^{(\textrm{b})}$ in Eq.~\eqref{eq:unitary-ebit} is written as
\begin{align}
\hat{\Pi}^{(\textrm{b})}:=&\frac{1}{2\abs{\mathcal{R}}}\sum_{(i, j):j-i\in\mathcal{R}}\hat{P}\left(\ket{i}_\textrm{B}-\ket{j}_\textrm{B}\right)\nonumber\\
\begin{split}
=&\frac{1}{2\abs{\mathcal{R}}}\sum_{(i, j):j-i\in\mathcal{R}}\left[\hat{P}\left(\ket{i}_\textrm{B}\right)+\hat{P}\left(\ket{j}_\textrm{B}\right)\right.\\
&\left.-\Bigl(\ket{i}_\textrm{B}{}_\textrm{B}\bra{j}+\ket{j}_\textrm{B}{}_\textrm{B}\bra{i}\Bigr)\right]
\end{split}\nonumber\\
=&\frac{1}{2\abs{\mathcal{R}}}\sum_{(m, n):\abs{m-n}\in\mathcal{R}}\left[\hat{P}\left(\ket{m}_\textrm{B}\right)-\ket{m}_\textrm{B}{}_\textrm{B}\bra{n}\right]\nonumber\\
=&\frac{1}{2}\hat{1}_\textrm{B}-\frac{1}{2\abs{\mathcal{R}}}\sum_{(m, n):\abs{m-n}\in\mathcal{R}}\ket{m}_\textrm{B}{}_\textrm{B}\bra{n},
\end{align}
which concludes Eq.~\eqref{eq:Pi}.

Finally, we detail transformation in Eq.~\eqref{eq:unitary-eph} as follows.
\begin{align}
&\hat{U}\hat{e}^{(\textrm{ph})}\hat{U}^\dagger\nonumber\\
\begin{split}
=&\frac{1}{2\abs{\mathcal{R}}}\sum_{(i, j):j-i\in\mathcal{R}}\left[\hat{P}(\ket{1}_{\textrm{A}, i})+\hat{P}(\ket{1}_{\textrm{A}, j})\right]\\
&\otimes\left[\hat{P}(\ket{i}_\textrm{B})+\hat{P}(\ket{j}_\textrm{B})\right]
\end{split}\nonumber\\
=&\frac{1}{2\abs{\mathcal{R}}}\sum_{(m, n):\abs{m-n}\in\mathcal{R}}\left[\hat{P}(\ket{1}_{\textrm{A}, m})+\hat{P}(\ket{1}_{\textrm{A}, n})\right]\otimes\hat{P}(\ket{m}_\textrm{B})\nonumber\\
=&\frac{1}{2\abs{\mathcal{R}}}\sum_{(m, n):\abs{m-n}\in\mathcal{R}}\sum_{\vec{a}}\hat{P}\left(\ket{\vec{a}}_\textrm{A}\right)\left(\delta_{a_m, 1}+\delta_{a_n, 1}\right)\otimes\hat{P}\left(\ket{m}_\textrm{B}\right)\nonumber\\
\begin{split}
=&\sum_{\vec{a}}\hat{P}\left(\ket{\vec{a}}_\textrm{A}\right)\\
&\otimes\frac{1}{2\abs{\mathcal{R}}}\sum_{m=1}^L\sum_{n:\abs{m-n}\in\mathcal{R}}\left(\delta_{a_m, 1}+\delta_{a_n, 1}\right)\hat{P}\left(\ket{m}_\textrm{B}\right)
\end{split}\nonumber\\
\begin{split}
=&\sum_{\vec{a}}\hat{P}\left(\ket{\vec{a}}_\textrm{A}\right)\\
&\otimes\sum_{m=1}^L\hat{P}\left(\ket{m}_\textrm{B}\right)\left(\frac{1}{2}\delta_{a_m, 1}+\frac{1}{2\abs{\mathcal{R}}}\sum_{n:\abs{m-n}\in\mathcal{R}}\delta_{a_n, 1}\right)
\end{split}\nonumber\\
=:&\sum_{\vec{a}}\hat{P}\left(\ket{\vec{a}}_\textrm{A}\right)\otimes\hat{\Pi}_{\vec{a}}^{(\textrm{ph})}.
\end{align}

\section{Proof of Lemma~\ref{lem:omegaplus-estimated}}
\label{append:lemma2}
Here, we prove Lemma~\ref{lem:omegaplus-estimated} in the main text. We consider the maximization of the largest eigenvalue of $\hat{P}_{\vec{a}}(\hat{\Pi}_{\vec{a}}^{(\textrm{ph})}-\lambda\hat{\Pi}^{(\textrm{b})})\hat{P}_{\vec{a}}$ in Eq.~\eqref{set:omegaplus} over $\vec{a}$ with $\abs{\vec{a}}=\nu+1$. By using Eq.~\eqref{eq:operator-omegaminus}, $\hat{P}_{\vec{a}}(\hat{\Pi}_{\vec{a}}^{(\textrm{ph})}-\lambda\hat{\Pi}^{(\textrm{b})})\hat{P}_{\vec{a}}$ is written as
\begin{align}
&\hat{P}_{\vec{a}}(\hat{\Pi}_{\vec{a}}^{(\textrm{ph})}-\lambda\hat{\Pi}^{(\textrm{b})})\hat{P}_{\vec{a}}\nonumber\\
\begin{split}
=&\sum_{m=1}^L\hat{P}\left(\ket{m}_\textrm{B}\right)\delta_{a_m, 1}\left(\frac{\delta_{a_m, 1}-\lambda}{2}+\frac{1}{2\abs{\mathcal{R}}}\sum_{n:\abs{m-n}\in\mathcal{R}}\delta_{a_n, 1}\right)\\
&+\frac{\lambda}{2\abs{\mathcal{R}}}\sum_{(m, n):\abs{m-n}\in\mathcal{R}}\ket{m}_\textrm{B}{}_\textrm{B}\bra{n}\delta_{a_m, 1}\delta_{a_n, 1}
\end{split}\nonumber\\
\begin{split}
=&\sum_{m: a_m=1}\left[\hat{P}\left(\ket{m}_\textrm{B}\right)\left(\frac{1-\lambda}{2}+\frac{1}{2\abs{\mathcal{R}}}\sum_{m:\abs{m-n}\in\mathcal{R}}\delta_{a_n, 1}\right)\right.\\
&\left.+\frac{\lambda}{2\abs{\mathcal{R}}}\sum_{m:\abs{m-n}\in\mathcal{R}}\ket{m}_\textrm{B}{}_\textrm{B}\bra{n}\delta_{a_n, 1}\right]
\end{split}\nonumber\\
\begin{split}
=&\sum_{m: a_m=1}\left\{\frac{1-\lambda}{2}\hat{P}\left(\ket{m}_\textrm{B}\right)\right.\\
&\left.+\sum_{n:\abs{m-n}\in\mathcal{R}\wedge a_n=1}\left[\frac{1}{2\abs{\mathcal{R}}}\hat{P}\left(\ket{m}_\textrm{B}\right)+\frac{\lambda}{2\abs{\mathcal{R}}}\ket{m}_\textrm{B}{}_\textrm{B}\bra{n}\right]\right\}
\end{split}\nonumber\\
=&\sum_{m: a_m=1}\frac{1-\lambda}{2}\hat{P}\left(\ket{m}_\textrm{B}\right)\nonumber\\
&+\sum_{\substack{(m, n):\\\abs{m-n}\in\mathcal{R}\wedge(a_m, a_n)=(1, 1)}}\left[\frac{1}{2\abs{\mathcal{R}}}\hat{P}\left(\ket{m}_\textrm{B}\right)+\frac{\lambda}{2\abs{\mathcal{R}}}\ket{m}_\textrm{B}{}_\textrm{B}\bra{n}\right].
\label{eq:operator-omegaplus}
\end{align}
Since $\mathcal{R}=\bigcup_{m=1}^t\{m, L-m\}$, the coefficient of $\hat{P}(\ket{m}_\textrm{B})$ for $m$ such that $a_m=1$ in Eq.~\eqref{eq:operator-omegaplus} is written as
\begin{align}
&\frac{1-\lambda}{2}+\frac{1}{2\abs{\mathcal{R}}}\#\left\{n\relmiddle|
\begin{array}{l}
n\in\{1, \dots, L\}\\
\wedge\abs{m-n}\in\mathcal{R}\wedge a_n=1
\end{array}
\right\}
\nonumber\\
=&\frac{1-\lambda}{2}+\frac{1}{2\abs{\mathcal{R}}}\sum_{n=1}^L\delta_{a_n, 1}\sum_{l=1}^t\left(\delta_{n, m+_Ll}+\delta_{n, m-_Ll}\right).
\end{align}
Here, $\#\{n\mid A(n)\}$ denotes the number of $n$ satisfying the condition $A(n)$, and $-_L$ denotes subtraction modulo $L$, namely, for integers $(p, q)$ with $1\leq p\leq L$ and $1\leq q \leq L$,
\begin{align}
p-_Lq=
\begin{cases}
p-q&\textrm{if }p\geq q+1,\\
p-q+L&\textrm{if }p\leq q.
\end{cases}
\end{align}
Also, the coefficient of $\ket{m}_\textrm{B}{}_\textrm{B}\bra{n}$ for $m, n$ such that $(a_m, a_n)=(1, 1)$ in Eq.~\eqref{eq:operator-omegaplus} is written as
\begin{align}
\begin{cases}
\frac{\lambda}{2\abs{\mathcal{R}}}&\textrm{if }n=m\pm_L l~(1\leq l\leq t),\\
0&\textrm{otherwise}.
\end{cases}
\end{align}
Next, we classify $\vec{a}$ in terms of the resulting eigenvalues of Eq.~\eqref{eq:operator-omegaplus}.
In so doing, note that the translation operation (as this is a unitary operator) defined by Eq.~\eqref{defi:translation-operator} does not change the eigenvalues of Eq.~\eqref{eq:operator-omegaplus}.
Hence, the eigenvalues of Eq.~\eqref{eq:operator-omegaplus} with $\vec{a}$ and $\vec{a}'$ are the same if there exists $\kappa$ ($1\leq \kappa\leq L$) such that $a_k=a'_{k+_L\kappa}$ is satisfied for any $k\in\{1,\dots, L\}$ and hence it is suffice to consider $\vec{a}\in[\vec{a}]=\{\vec{a}'\mid\exists\kappa\in\{1,\dots, L\}\textrm{ s.t. }a'_{k+_L\kappa}=a_k\textrm{ for }\forall k\in\{1,\dots, L\}\}$ to derive the eigenvalues of Eq.~\eqref{eq:operator-omegaplus}.
In order to characterize $\vec{a}$, we introduce an $\abs{\vec{a}}(=\nu+1)$-length vector $\vec{p}=(p_1p_2\dots p_{\nu+1})$ that satisfies
\begin{align}
a_{p_j}=1\textrm{ for }\forall j\in\{1,\dots,\nu+1\},\\
1\leq p_1<p_2<\dots<p_{\nu+1}\leq L.
\end{align}
By using $\vec{p}$, we can convert the problem of deriving the largest eigenvalue of Eq.~\eqref{eq:operator-omegaplus} to the maximization problem of the largest eigenvalue of the following matrix.
\begin{align}
\frac{1-\lambda}{2}I_{\nu+1}+\frac{1}{2\abs{\mathcal{R}}}B_{\nu+1}(\lambda).\label{eq:matrix-omegaplus-general}
\end{align}
Here, $I_{\nu+1}$ denotes the $(\nu+1)\times(\nu+1)$ identity matrix and $B_{\nu+1}(\lambda)$ denotes the $(\nu+1)\times(\nu+1)$ matrix whose diagonal element $\left(B_{\nu+1}(\lambda)\right)_{m, m}$ is given by
\begin{align}
\left(B_{\nu+1}(\lambda)\right)_{m, m}=&\#\{k\mid k\in\{1, \dots, \nu+1\}\wedge\abs{p_m-p_k}\in\mathcal{R}\}\nonumber\\
=&\sum_{k=1}^{\nu+1}\sum_{l=1}^{t}\left(\delta_{p_k, p_m+_Ll}+\delta_{p_k, p_m-_Ll}\right)\nonumber\\
=&\sum_{l=1}^{t}\sum_{k=1}^{\nu+1}\left(\delta_{p_k, p_m+_Ll}+\delta_{p_k, p_m-_Ll}\right)\nonumber\\
=&\sum_{l=1}^{t}\sum_{k=1:k\neq m}^{\nu+1}\left(\delta_{p_k, p_m+_Ll}+\delta_{p_k, p_m-_Ll}\right),
\end{align}
and its off-diagonal element $\left(B_{\nu+1}(\lambda)\right)_{m, n}$ ($m\neq n$) is given by
\begin{align}
\left(B_{\nu+1}(\lambda)\right)_{m, n}=
\begin{cases}
\lambda&\textrm{if }p_n=p_m\pm_Ll~(1\leq l\leq t).\\
0&\textrm{otherwise}.
\end{cases}
\end{align}
Among $[\vec{a}]$, we need to find $\vec{a}$ that achieves the largest eigenvalue of Eq.~\eqref{eq:matrix-omegaplus-general}.
For this, we use the following fact.
\begin{fact}
For any real matrix with non-negative off-diagonal elements, the largest eigenvalue is maximized when all the matrix elements are maximized.
\label{fact:maximized-matrix-form}
\end{fact}
\textit{Proof.}
We consider two $n\times n$ real matrices $A=(A_{i, j})_{i, j}$ and $\tilde{A}=(\tilde{A}_{i, j})_{i, j}$ such that $A_{i, j}\geq\tilde{A}_{i,j}$ holds for any $i, j\in\{1,\dots n\}$ and $A_{i, j}, \tilde{A}_{i, j}\geq0$ holds if $i\neq j$.
Suppose that $\ket{\psi}=(x_1 x_2 \dots x_n)^\mathrm{T}$ and $\ket{\tilde{\psi}}=(\tilde{x}_1 \tilde{x}_2 \dots \tilde{x}_n)^\mathrm{T}$ are normalized eigenvectors of $A$ and $\tilde{A}$ that give the largest eigenvalue of $A$ and $\tilde{A}$, respectively. Since both $A$ and $\tilde{A}$ are real and all the off-diagonal elements of $A$ and $\tilde{A}$ are non-negative, we can choose $\ket{\psi}$ and $\ket{\tilde{\psi}}$ such that all the elements of $\ket{\psi}$ and $\ket{\tilde{\psi}}$ are real and non-negative. By using $\ket{\psi}$ and $\ket{\tilde{\psi}}$, the largest eigenvalue of $A$ and $\tilde{A}$ are respectively given by
\begin{align}
\braket{A}_{\max}:=\bra{\psi}A\ket{\psi}=\sum_{i, j}A_{i, j}x_ix_j,\\
\braket{\tilde{A}}_{\max}:=\bra{\tilde{\psi}}\tilde{A}\ket{\tilde{\psi}}=\sum_{i, j}\tilde{A}_{i, j}\tilde{x}_i\tilde{x}_j.
\end{align}
Since $A_{i, j}\geq\tilde{A}_{i,j}$ holds for any $i, j\in\{1,\dots n\}$ and $\ket{\psi}$ gives the largest eigenvalue of $A$, we have
\begin{align}
\braket{\tilde{A}}_{\max}=&\sum_{i, j}\tilde{A}_{i, j}\tilde{x}_i\tilde{x}_j\leq\sum_{i, j}A_{i, j}\tilde{x}_i\tilde{x}_j=\bra{\tilde{\psi}}A\ket{\tilde{\psi}}\nonumber\\
\leq&\bra{\psi}A\ket{\psi}=\braket{A}_{\max},
\end{align}
which ends the proof.

By using Fact~\ref{fact:maximized-matrix-form}, the largest eigenvalue of Eq.~\eqref{eq:matrix-omegaplus-general} is obtained when $p_j+1=p_{j+1}$ for all $j\in\{1,\dots,\nu\}$, namely, $\vec{a}=(0\dots0\overbrace{1\dots1}^{\nu+1}0\dots0)$ in Eq.~\eqref{eq:operator-omegaplus}.
For example, if $\nu=2$ and $\abs{\mathcal{R}}=2$, Eq.~\eqref{eq:matrix-omegaplus-general} with $\vec{a}=(0\dots01110\dots0)$ is rewritten as
\begin{align}
\begin{bmatrix}
\frac{1-\lambda}{2}+\frac{1}{4}&\frac{\lambda}{4}&0\\
\frac{\lambda}{4}&\frac{1-\lambda}{2}+\frac{1}{2}&\frac{\lambda}{4}\\
0&\frac{\lambda}{4}&\frac{1-\lambda}{2}+\frac{1}{4}
\end{bmatrix},
\end{align}
and this results in the largest eigenvalue of Eq.~\eqref{eq:operator-omegaplus}, which corresponds to $\Omega_+^{(2)}(\lambda)$ for $\mathcal{R}=\{1, L\}$.
Moreover, if $\nu\leq t~(=\abs{\mathcal{R}}/2)$, Eq.~\eqref{eq:matrix-omegaplus-general} with $\vec{a}=(0\dots0\overbrace{1\dots1}^{\nu+1}0\dots0)$ is rewritten as
\begin{align}
\frac{1-\lambda}{2}I_{\nu+1}+\frac{1}{2\abs{\mathcal{R}}}\mathcal{B}_{\nu+1}(\lambda),\label{eq:matrix-omegaplus}
\end{align}
where $\mathcal{B}_{\nu+1}(\lambda)$ denotes the $(\nu+1)\times(\nu+1)$ matrix whose elements are given by
\begin{align}
\left(\mathcal{B}_{\nu+1}(\lambda)\right)_{m, n}=
\begin{cases}
\nu&\textrm{if }m=n,\\
\lambda&\textrm{otherwise}.
\end{cases}
\end{align}
Eq.~\eqref{eq:matrix-omegaplus} has only two eigenvalues: $(1-\lambda)/2+(\nu-\lambda)/(2\abs{\mathcal{R}})$ and $(1-\lambda)/2+\nu(1+\lambda)/(2\abs{\mathcal{R}})$. Since $\lambda\geq0$, we have
\begin{align}
\Omega_+^{(\nu)}(\lambda)=\frac{1-\lambda}{2}+\nu\frac{1+\lambda}{2\abs{\mathcal{R}}},
\end{align}
which concludes Eq.~\eqref{eq:omegaplus-estimated}.

\section{Proof of Theorem~\ref{theo:phase_error_nu=1}}
\label{append:theorem1}
Here, we prove Theorem~\ref{theo:phase_error_nu=1} in the main text. In order to maximize $e_-^{(\textrm{ph}, 1)}$, we derive an upper bound on $\Omega_-^{(1)}{(\lambda)}$. For this, we consider the largest eigenvalue of $\hat{\Pi}_{\vec{a}}^{(\textrm{ph})}-\lambda\hat{\Pi}^{(\textrm{b})}$ for $\abs{\vec{a}}=\nu-1=0$, namely, $\vec{a}=(00\dots0)=:\vec{0}$. Since $\hat{\Pi}_{\vec{0}}^{(\textrm{ph})}=0$, $\Omega_-^{(1)}(\lambda)$ is given by the largest eigenvalue of $-\lambda\hat{\Pi}^{(\textrm{b})}$, which is non-positive. Hence, $e_-^{(\textrm{ph}, 1)}$ is upper bounded by $e_-^{(\textrm{ph}, 1)}=\min_{\lambda\geq0}\{\lambda e^{(\textrm{b}, 1)}+\Omega_-^{(1)}(\lambda)\}\leq\min_{\lambda\geq0}\{\lambda e^{(\textrm{b}, 1)}\}=0$.
For $e_+^{(\textrm{b}, 1)}$, from Eq.~\eqref{ineq:ephplusminus-bound-detail}, we obtain
\begin{align}
e_+^{(\textrm{ph}, 1)}\leq\frac{1}{2}+\frac{1}{2\abs{\mathcal{R}}}\label{ineq:ephbound-nu=1-higherror}
\end{align}
for $e^{(\textrm{b}, 1)}\geq\frac{\abs{\mathcal{R}}-1}{2\abs{\mathcal{R}}}$. Also, for $0\leq e^{(\textrm{b}, 1)}\leq\frac{\abs{\mathcal{R}}-1}{2\abs{\mathcal{R}}}$, $e^{(\textrm{ph}, 1)}$ is upper bounded by
\begin{align}
e^{(\textrm{ph}, 1)}\leq\frac{2\abs{\mathcal{R}}}{\abs{\mathcal{R}}-1}e^{(\textrm{b}, 1)}\left(\frac{1}{2}+\frac{1}{2\abs{\mathcal{R}}}\right)=\frac{\abs{\mathcal{R}}+1}{\abs{\mathcal{R}}-1}e^{(\textrm{b}, 1)}\label{ineq:ephbound-nu=1-lowerror}
\end{align}
by choosing $p$ in Eq.~\eqref{ineq:eph-mixture} as $p=\frac{2\abs{\mathcal{R}}}{\abs{\mathcal{R}}-1}e^{(\textrm{b}, 1)}$. Therefore, by combining Eqs.~\eqref{ineq:ephbound-nu=1-higherror} and~\eqref{ineq:ephbound-nu=1-lowerror}, we conclude Eq.~\eqref{eq:phase_error_nu=1}.


\end{document}